\documentclass[twocolumn,epjc3,envcountsect]{svjour3}

\RequirePackage[T1]{fontenc}
\usepackage[american]{babel}
\usepackage[normalem]{ulem}

\usepackage{comment}

\RequirePackage[T1]{fontenc}

\smartqed

\RequirePackage{graphicx}
\RequirePackage{mathptmx}
\RequirePackage{flushend}
\RequirePackage[numbers,sort&compress]{natbib}
\RequirePackage[colorlinks,citecolor=blue,urlcolor=blue,linkcolor=blue]{hyperref}

\usepackage{amssymb}
\usepackage{color,colortbl}
\usepackage{epstopdf}
\usepackage{graphicx}
\usepackage{tabularx}
\usepackage{siunitx} 
\usepackage{booktabs}
\usepackage{amsmath}
\usepackage{dsfont}
\usepackage{xstring}
\usepackage{subfigure}

\usepackage{microtype} 

\usepackage{arydshln}

\usepackage[capitalize]{cleveref}
\Crefname{section}{Section}{Sections}
\Crefname{figure}{Figure}{Figures}
\Crefname{table}{Table}{Tables}
\Crefname{appendix}{Appendix}{Appendices}
\Crefname{equation}{Eq.}{Eqs.}
\newcommand{\Onlinecite}[1]{%
    \IfSubStr{#1}{,}{Refs}{Ref}.~\cite{#1}%
}

\DeclareMathOperator{\tr}{\rm tr\,}

\definecolor{Gray}{gray}{0.9}

\journalname{Eur. Phys. J. C}

\begin{document}

\title{Unveiling the flux tube structure in full QCD}

\author{M. Baker\thanksref{e1,addr1}
\and
P. Cea\thanksref{e2,addr2}
\and
V. Chelnokov\thanksref{e3,addr3}
\and
L. Cosmai\thanksref{e4,addr2}
\and
A. Papa\thanksref{e5,addr4,addr5}
}

\institute{Department of Physics, University of Washington, WA 98105 Seattle, USA\label{addr1}
\and
INFN - Sezione di Bari, I-70126 Bari, Italy\label{addr2}
\and
Institut f\"ur Theoretische Physik, Goethe Universit\"at, 60438 Frankfurt am Main, Germany\label{addr3}
\and
Dipartimento di Fisica dell'Universit\`a della Calabria, I-87036 Arcavacata di Rende, Cosenza, Italy\label{addr4}
\and
INFN - Gruppo collegato di Cosenza, I-87036 Arcavacata di Rende, Cosenza, Italy\label{addr5}
}

\thankstext{e1}{e-mail: mbaker4@uw.edu}
\thankstext{e2}{e-mail: paolo.cea@ba.infn.it}
\thankstext{e3}{e-mail: chelnokov@itp.uni-frankfurt.de}
\thankstext{e4}{e-mail: leonardo.cosmai@ba.infn.it}
\thankstext{e5}{e-mail: alessandro.papa@fis.unical.it}

\date{Received: date / Accepted: date}

\maketitle

\begin{abstract}
We present lattice Monte Carlo results on the chromoelectric field created by a static quark-antiquark pair in the vacuum of QCD with 2+1 dynamical staggered fermions at physical masses. After isolating the nonperturbative, confining part of the field, we characterize its spatial profile for several values of the physical distances between the sources, ranging from about 0.5~fm up to the onset of string breaking. Moreover, we compare our results with a model of QCD vacuum as disordered chromomagnetic condensate.
\end{abstract}

\section{Introduction}
Aiming at a deeper understanding of the microscopic mechanisms of confinement, in the recent years we have investigated, by numerical Monte Carlo simulations of the SU(3) pure gauge theory on a space-time lattice, the detailed structure of the color fields in the region around two static sources, a quark and an antiquark, both at zero~\cite{Baker:2018mhw,Baker:2019gsi,Baker:2022cwb} and nonzero temperature~\cite{Baker:2023dnn}. Some of our numerical results have been considered within phenomenological models for hadronization (see, {\it e.g.}, Refs.~\cite{Bierlich:2022oja,Bierlich:2020naj}).

The following key facts have emerged:
\begin{itemize}
    \item The (chromo-)magnetic field $\vec{B}$ around the sources is compatible with zero within uncertainties. The components of the (chromo-)electric field $\vec{E}$ transverse to the line connecting the sources consist solely of a perturbative, irrotational, short-distance contribution~\footnote{From now on, we omit the prefix "chromo-" for fields.}. The longitudinal component of $\vec{E}$ can be separated into the perturbative, short-distance part and a  nonperturbative term $\vec{E}^{NP}$, encoding the confining information, which is shaped as a smooth {\em flux tube}.

    \item The curl of the  electric field determines a magnetic current $\vec{J}_{\rm mag}$ which circulates about the axis of the flux tube and exhibits continuum scaling. The force on the magnetic currents in flux tubes has the  form $\vec{f} = \vec{J}_\text{\rm mag} \times \vec{E}^\text{NP}$,   where $\vec{J}_\text{mag}$ and $\vec{E}^\text{NP}$ are measured 
 by lattice simulations, thus supporting the Maxwell picture of confinement  ~\cite{Baker:2019gsi}. (See also  the theoretical analysis of Ref.~\cite{Cea:2023} which states that the longitudinal electric field is dominated by its Abelian components.)
\end{itemize}

The purpose of the present work is to probe the Maxwell scenario in the case of full QCD with dynamical quarks. Here we expect it to be valid in a manner somewhat similar to the SU(3) pure gauge, as long as the distance between the static sources does not reach the value at which {\em string breaking}~\cite{Philipsen:1998de,Kratochvila:2002vm,Bali:2005fu,Koch:2015qxr} occurs and the flux-tube structure disappears.
At the onset of the string breaking phenomenon, however, some new features in the behavior of color fields and currents must appear, which can be detected in our numerical simulations.  

The paper is organized as follows: in Section~2, we recall the theoretical background of our approach and the Maxwell picture of confinement; in Section~3, we present the setup of our numerical simulations (lattice action, scale setting, smearing procedure); in Section~4, we show our results; in Section~5, we present the basic formulas derived from the model of Ref.~\cite{Cea:2023} to be fitted to numerical data; in Section~6, we discuss possible evidence for string breaking in our numerical data; finally, in Section~7, we draw our conclusions.

\section{Connected correlator, field strength tensor, Maxwell picture of the flux tube}
\begin {figure}[htb]
\centering
\includegraphics[width=0.7\linewidth,clip]{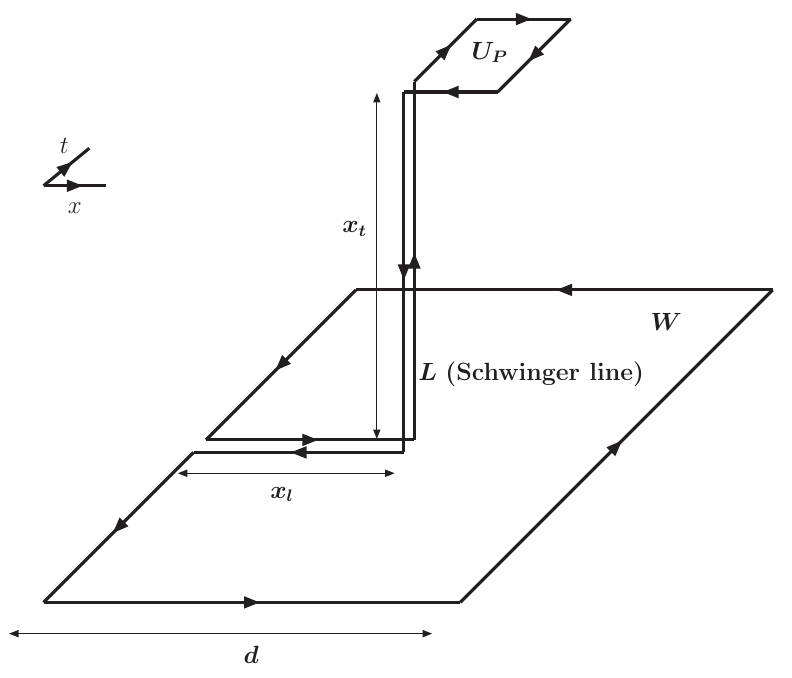}
\includegraphics[width=0.25\textwidth,clip]{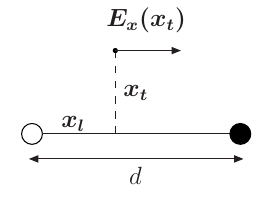}
\caption{The connected correlator between the plaquette $U_{P}$ and the Wilson loop (subtraction in $\rho_{W,\,\mu\nu}^\text{conn}$ not explicitly drawn).
The longitudinal electric field $E_x(x_t)$ at some fixed displacement $x_l$ along the axis connecting the static sources (represented by the white and black circles), for a given value of the transverse distance $x_t$.}               
\label{fig:op_W}
\end{figure}

 Similarly to our previous studies in $\text{SU(3)}$ pure gauge theory~\cite{Baker:2018mhw,Baker:2019gsi,Baker:2022cwb}, we determine the spatial distributions of the color fields induced by a static quark-antiquark pair from lattice measurements of the connected correlation function $\rho^\text{conn}_{W, \mu \nu}$~\cite{DiGiacomo:1989yp} of a plaquette $U_P = U_{\mu \nu} (x)$ in the ${\mu \nu}$ plane, and a square Wilson loop $W$ (see Fig.~\ref{fig:op_W}),
\begin{equation}
    \rho^\text{conn}_{W, \mu \nu} = \frac {\langle\tr (WLU_PL^\dagger)\rangle}{\langle\tr(W)\rangle} - \frac{1}{N} \frac {\langle\tr (U_P) \tr (W)\rangle}{\langle\tr(W)\rangle}\;,
    \label{connected1}
\end{equation}
$N=3$ being the number of QCD colors. The correlator $\rho^\text{conn}_{W, \mu \nu}$ provides a lattice definition of a gauge-invariant field strength tensor $\langle F_{\mu \nu}\rangle_{q \bar{q}} \equiv F_{\mu \nu}$ carrying a unit of octet charge, while possessing the space-time symmetry properties of the Maxwell field tensor of electrodynamics, 
\begin{equation}
\rho^\text{conn}_{W, \mu \nu} \equiv~~ a^2 g\langle F_{\mu \nu}\rangle_{q \bar{q}} ~~ \equiv~~ a^2 g ~F_{\mu \nu}\;.
\label{connected2}
\end{equation}

When the plaquette $U_P$ lies in the $\hat 4 \hat 1$ plane, the measured $\hat 4 \hat 1$ component of the field tensor determines $E_x$, the component of the electric field along the $q\bar{q}$ axis $ E_x = F_{41}$; {\it i.e.}, the longitudinal component of the electric field at the position corresponding to the center of the plaquette. 

When $U_P$ is in the $\hat 4 \hat 2$ plane, $F_{42} = E_y$, a component of the electric field transverse to the $q\bar{q}$ axis,
when $U_P$ is in the $\hat 2 \hat 3$ plane, $F_{23} = B_x$,  the  longitudinal component of the magnetic field, {\it etc}.

The Maxwell-like field $F_{\mu \nu}$, see Eq.~\eqref{connected2}, generates a magnetic current density $J_\alpha^\text{mag}$ that circulates about the axis of the flux tube: 

\begin{equation}
     J_\alpha^{\rm mag} \equiv \frac{1}{2} \epsilon_{\alpha \beta\mu \lambda} \frac{\partial F_{\mu \lambda}}{\partial x^\beta},~~~~~~~( \epsilon_{4123} = 1) \;.
     \label{Jmag}
\end{equation} 
(The presence of magnetic currents in SU(3) lattice gauge theory was already suggested in Ref.~\cite{Skala:1996ar}.)

The Maxwell stress tensor $T_{\alpha \beta}$ determines the force density $f_\beta$ acting on these currents:

\begin{equation}
 T_{\alpha \beta} = F_{\alpha \lambda} F_{\beta \lambda} - \frac{1}{4} \delta_{\alpha \beta} F_{\mu \lambda} F_{\mu \lambda}\ , 
 \label{stress}   
\end{equation}

\begin{equation}
    f_\beta =  \frac{\partial}{\partial x^\alpha} T_{\alpha \beta}\ =  - F_{\mu \lambda} ~\frac{1}{2} \epsilon_{\alpha \beta\mu \lambda} J^\text{mag}_\alpha \  .
    \label{Fs}
\end{equation}

 Under the assumption that the magnetic components of the field tensor, $\frac{1}{2} \epsilon_{ijk} F_{jk}$, vanish,  the spatial part of   Eq.~(\ref{Jmag})   becomes
\begin{equation}
\label{rotel}
  \vec{J}_\text{mag} = \vec{\nabla} \times \vec{E} \;,
 \end{equation} 
while the spatial part of  Eq.~(\ref{Fs})   becomes   $\vec{f} =  \vec{J}_\text{mag}  \times \vec{E}\ $.

Replacing the color electric field $\vec{E}$
by its   nonperturbative longitudinal component $\vec{E}^\text{NP}$, determined from our lattice simulations of the connected correlator, we obtain 
\begin{equation}   
  \vec{f} = \vec{J}_{\rm mag}  \times \vec{E}^\text{NP},
\label{vecfdensity}
\end{equation}
 the confining force density  directed toward the flux-tube axis.

The 'Maxwell' picture of confinement is supported by a recent analysis of the confinement in SU(3) theories~\cite{Cea:2023}, showing that the electric field $g E^a$ in the flux tube is mainly composed of the Abelian components $g E^3$ and $g E^8$.
\begin{table*}[th]
\begin{center} 
  \caption{Summary of the numerical simulations.}
  \label{numsimul}
\setlength{\tabcolsep}{20pt}
\begin{tabular}{cllclc}
\toprule
lattice & $\beta=10/g^2$ & $a(\beta)$ [fm]  & $d$ [lattice units]   & $d$ [fm] & statistics\\ \midrule
$	48^4	$	&	6.885	&	0.0949777	 &	6	&	0.569866	&	500	        \\
$	32^4	$	&	7.158	&	0.0738309	 &	8	&	0.590647	&	10064 	\\
$	24^4	$	&	6.445	&	0.144692	         &	5	&	0.723462	&	3330  	\\
$	32^4	$	&	7.158	&	0.0738309	 &	10	&	0.738309	&	10181	\\
$	48^4	$	&	6.885	&	0.0949777	 &	8	&	0.75982	&	779	        \\
$	32^4	$	&	6.885	&	0.0949777	 &	8	&	0.759821	&	4409	        \\
$	32^4	$	&	6.5824	&	0.126658	         &	6	&	0.759947	&	2667	       \\
$	32^4	$	&	6.3942	&	0.15203	         &	5	&	0.760151	&	3000	       \\
$	32^4	$	&	6.885	&	0.0949777	 &	9	&	0.854799	&	4347	       \\
$	32^4	$	&	6.25765	&	0.173715	         &	5	&	0.868573	&	3545	       \\
$	32^4	$	&	6.5824	&	0.126658	         &	7	&	0.886605	&	2667	      \\
$	32^4	$	&	6.3942	&	0.15203	         &	6	&	0.912182	&	3000	      \\
$	48^4	$	&	6.885	&	0.0949777	 &	10	&	0.949777	&	779	      \\
$	32^4	$	&	7.158	&	0.0738309	 &	13	&	0.959801	&	10183	\\
$	24^4	$	&	6.445	&	0.144692	         &	7	&	1.01285	&	3330  	\\
$	32^4	$	&	6.5824	&	0.126658	         &	8	&	1.01326	&	2666	       \\
$	32^4	$	&	7.158	&	0.0738309	 &	14	&	1.03363	&	2107	       \\
$	32^4	$	&	6.25765	&	0.173715	         &	6	&	1.04229	&	3549	      \\
$	32^4	$	&	6.885	&	0.0949777	 &	11	&	1.04475	&	4408	      \\
$	32^4	$	&	6.3942	&	0.15203	         &	7	&	1.06421	&	3000	      \\
$	32^4	$	&	6.33727	&	0.160714	         &	7	&	1.125	&	3133	      \\
$	32^4	$	&	6.885	&	0.0949777	 &	12	&	1.13973	&	4409	      \\
$	48^4	$	&	6.885	&	0.0949777	 &	12	&	1.13973	&	769	      \\
$	32^4	$	&	6.5824	&	0.126658	         &	9	&	1.13992	&	2667	      \\
$	32^4	$	&	6.314762	&	0.164286	         &	7	&	1.15	         &	3651 	\\
$	24^4	$	&	6.445	&	0.144692	         &	8	&	1.157536	&	3330	      \\
$	32^4	$	&	6.28581	&	0.168999	         &	7	&	1.18299	&	3148	     \\
$	32^4	$	&	6.25765	&	0.173715	         &	7	&	1.216	&	3546	     \\
$	32^4	$	&	6.3942	&	0.15203	         &	8	&	1.21624	&	3000	     \\
$	32^4	$	&	6.885	&	0.0949777	&	13	&	1.23471	&	4409	     \\
$	32^4	$	&	6.5824	&	0.126658     	&	10	&	1.26658	&	2667	     \\
$	32^4	$	&	6.3942	&	0.15203	         &	9	&	1.36827	&	3000	     \\
\bottomrule 
\end{tabular}
\end{center}
\end{table*}
\begin {figure}[thb]
\centering
\includegraphics[width=\linewidth,clip]{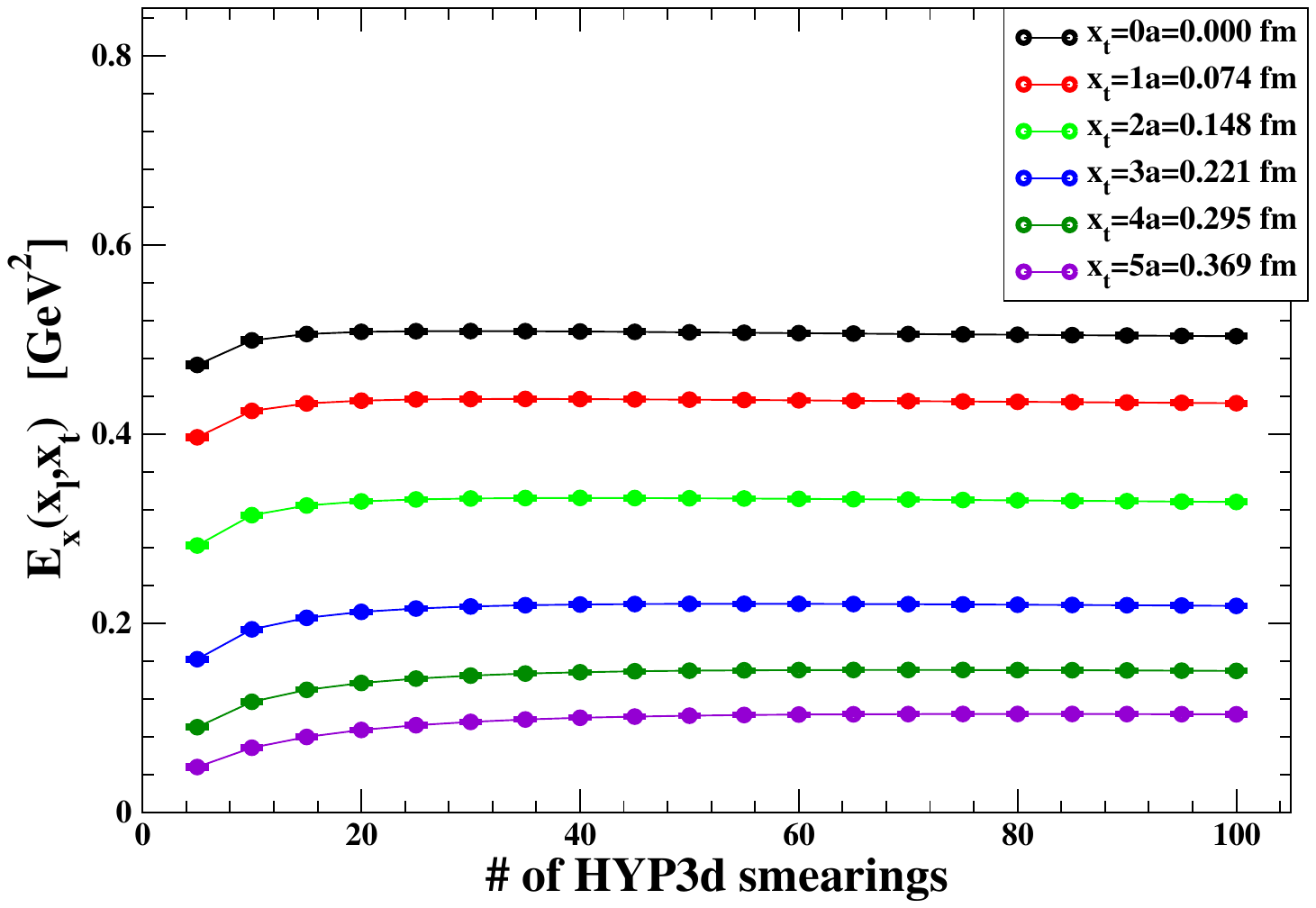}
\caption{Behavior under smearing of the "full" longitudinal electric field, $E_x$, for different values of the transverse distance $x_t$, at $\beta=7.158$ and $d=10a=0.74$ fm on a lattice $32^4$.}               
\label{fig:example_of_smearing}
\end{figure}
\section{Lattice setup and smearing procedure}
We perform simulations of lattice QCD with 2+1 flavors of HISQ (Highly Improved Staggered Quarks) 
quarks. We have made use of the HISQ/tree action~\cite{Follana:2006rc,Bazavov:2009bb,Bazavov:2010ru}.
Couplings are adjusted so as to move on a line of constant physics (LCP), as determined in Ref.~\cite{Bazavov:2011nk}, with the strange quark mass $m_s$ fixed at its physical value and a light-to-strange mass ratio $m_l/m_s=1/20$, corresponding to a pion mass of 160 MeV in the continuum limit.
We have simulated the theory for several values of 
the gauge coupling, adopting lattices of size $24^4$, $32^4$ and $48^4$. The thermalized lattice configurations, stored for further measurements, are each separated by 25 trajectories of rational hybrid Monte Carlo (RHMC) with length one.
The electromagnetic field tensor has been measured for a static quark and antiquark placed at a given distance $d$ in lattice spacing.
We fix the lattice spacing through the  observable $r_1$ as defined in the Appendix~B of  Ref.~\cite{Bazavov:2011nk}.

For the  $r_1$ scale  the lattice spacing
is given in terms of the $r_1$ parameter as:
\begin{equation}
\label{scale-r1}
\frac{a}{r_1}(\beta)_{m_l=0.05m_s}=
\frac{c_0 f(\beta)+c_2 (10/\beta) f^3(\beta)}{
1+d_2 (10/\beta) f^2(\beta)} \; ,
\end{equation}
with $c_0=44.06$, $c_2=272102$, $d_2=4281$, $r_1=0.3106(20)$ fm. \\
\begin{figure*}[htb]
\centering
\includegraphics[width=0.45\linewidth,clip]{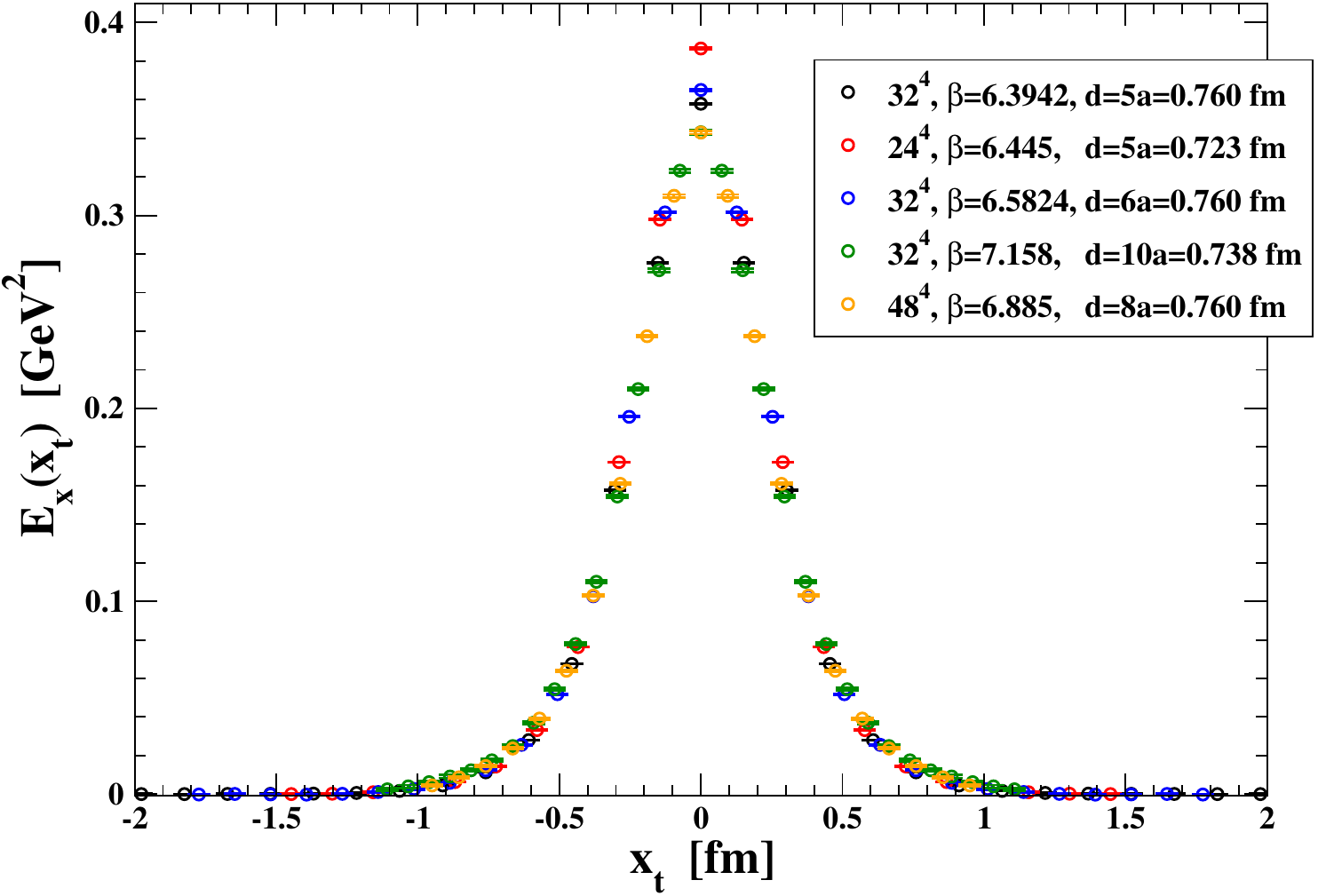}
\includegraphics[width=0.45\linewidth,clip]{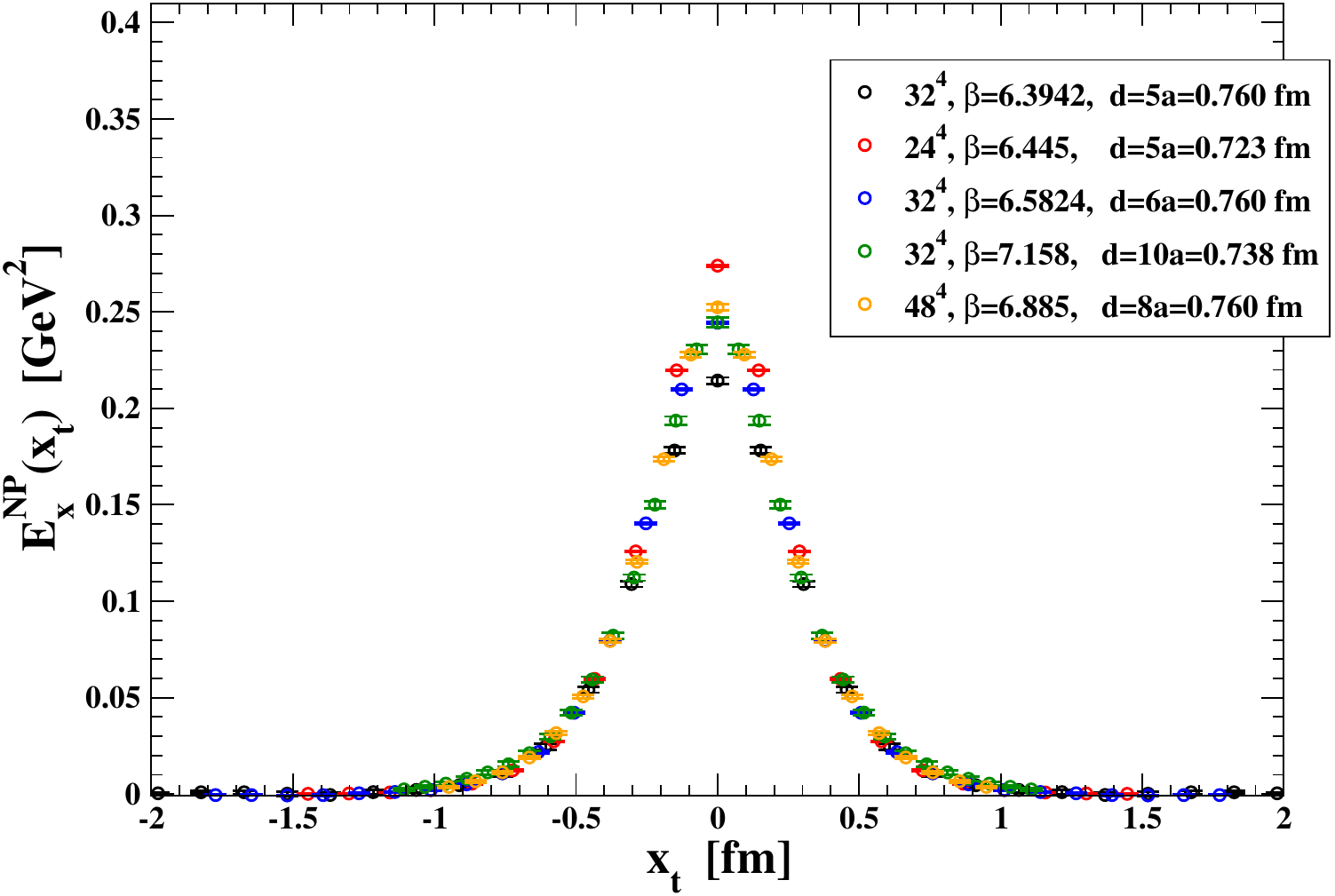}
\\
\includegraphics[width=0.45\linewidth,clip]{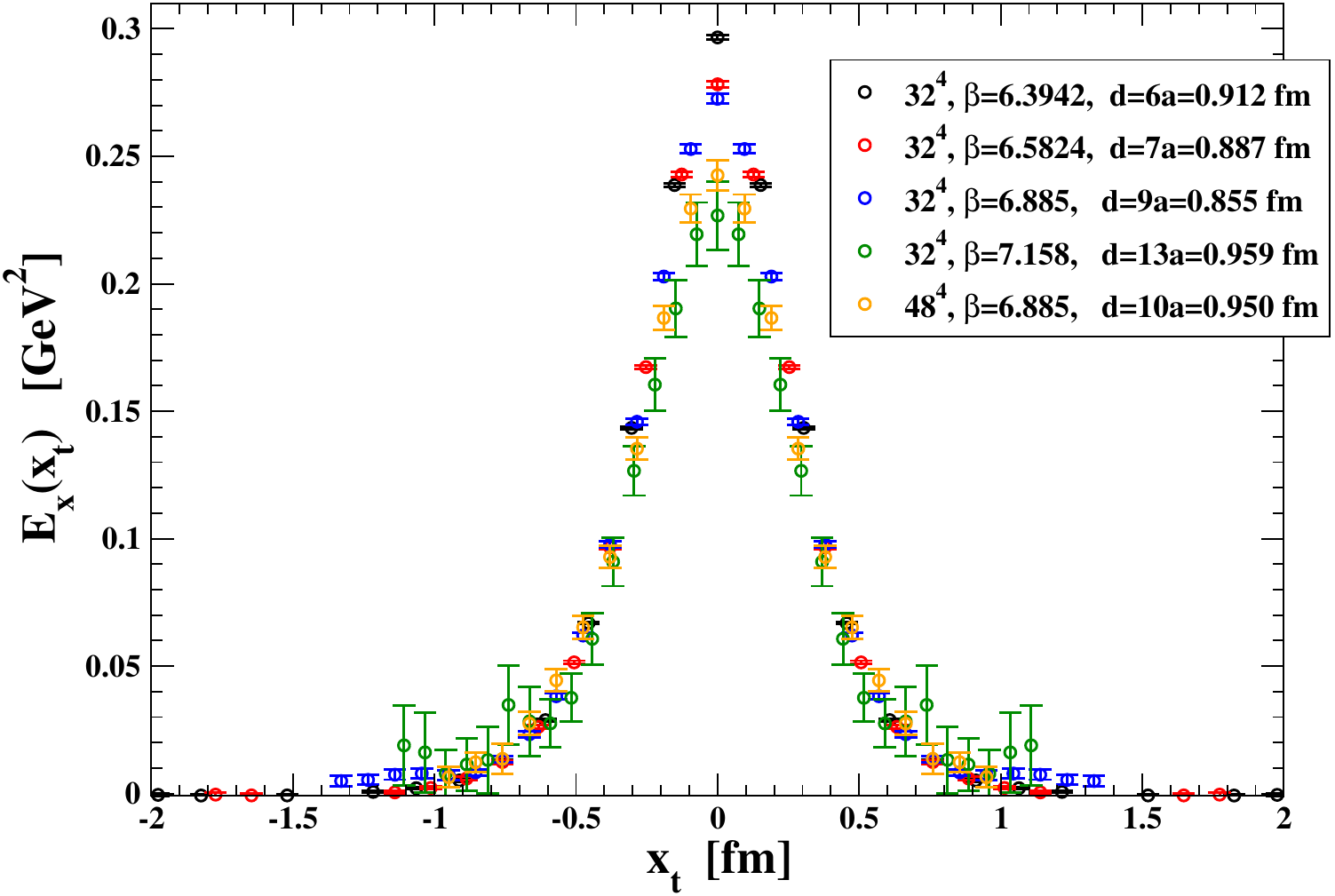}
\includegraphics[width=0.45\linewidth,clip]{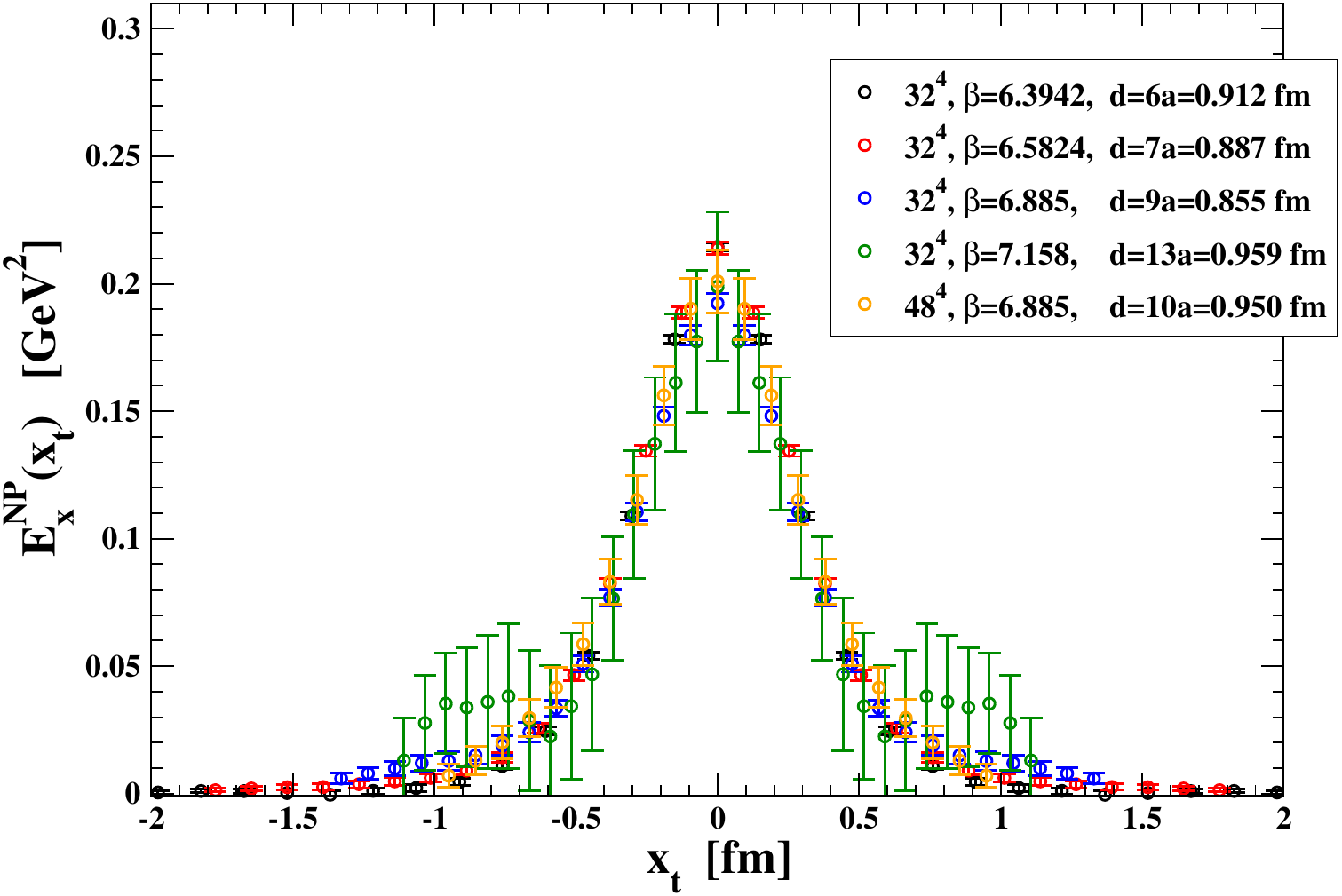}
\\
\includegraphics[width=0.45\linewidth,clip]{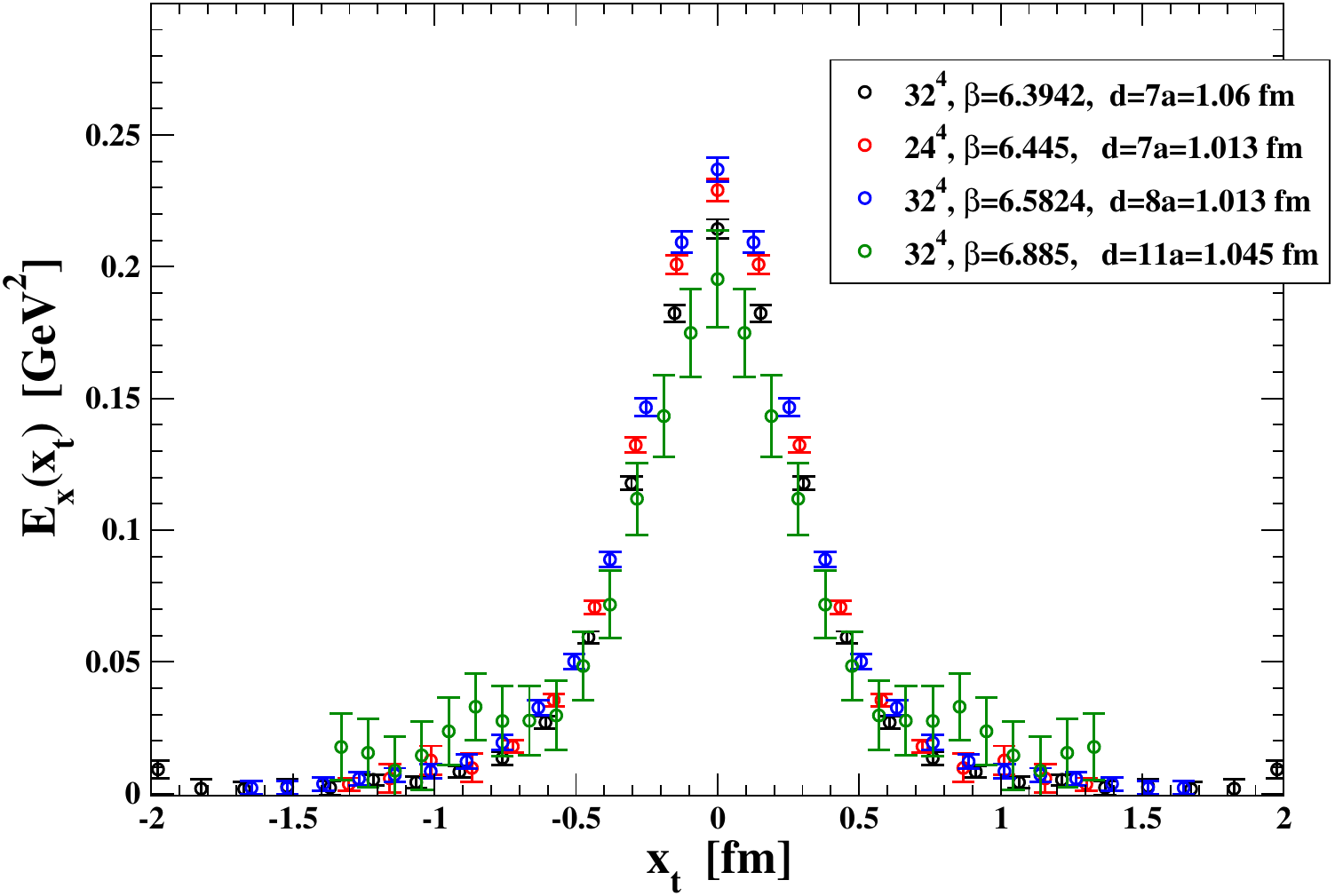}
\includegraphics[width=0.45\linewidth,clip]{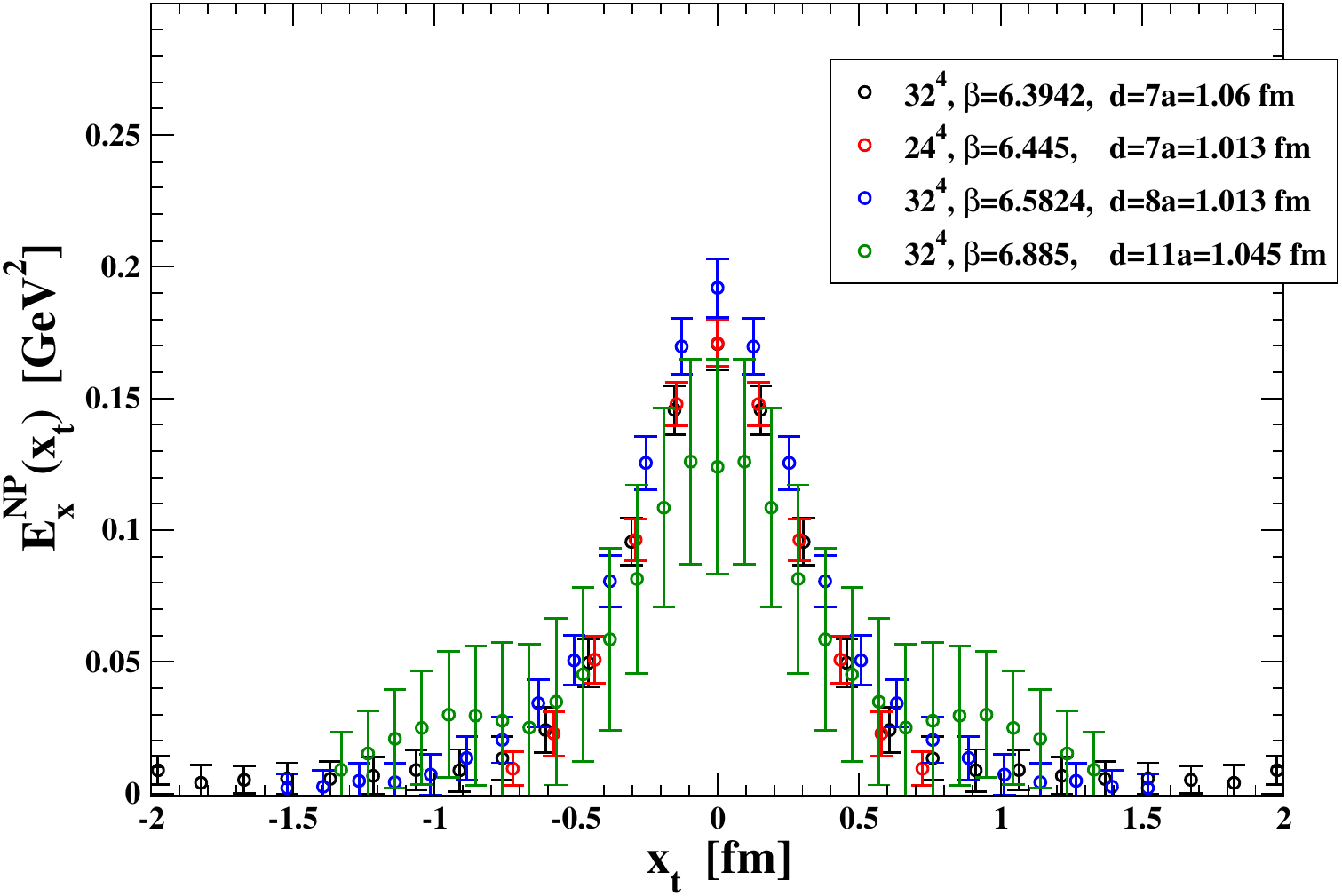}
\caption{Scaling test for the full electric field (left panels) and for its nonperturbative part (right panels) on the midplane for distances in the range $d=0.723-0.760$ fm (upper panels), $d=0.855-0.959$ fm (middle panels) and $d=1.013-1.060$ fm (lower panels).
}            
\label{fig:scaling}
\end{figure*}
\begin{figure*}[htb]
\vspace{-0.0cm}
\centering
\includegraphics[width=0.30\linewidth,clip]{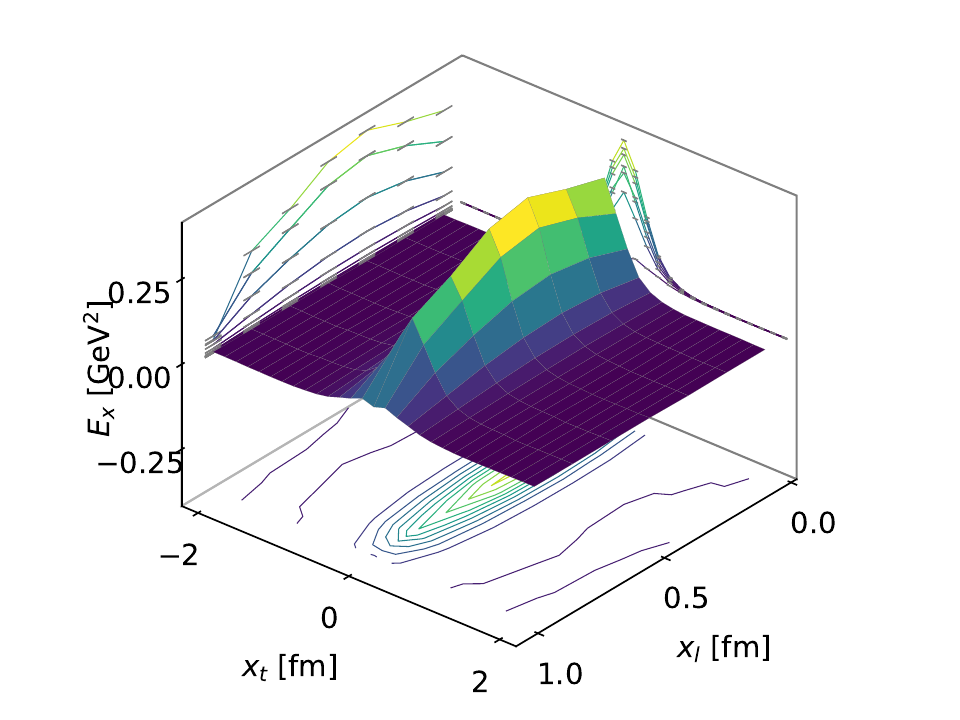}
\includegraphics[width=0.30\linewidth,clip]{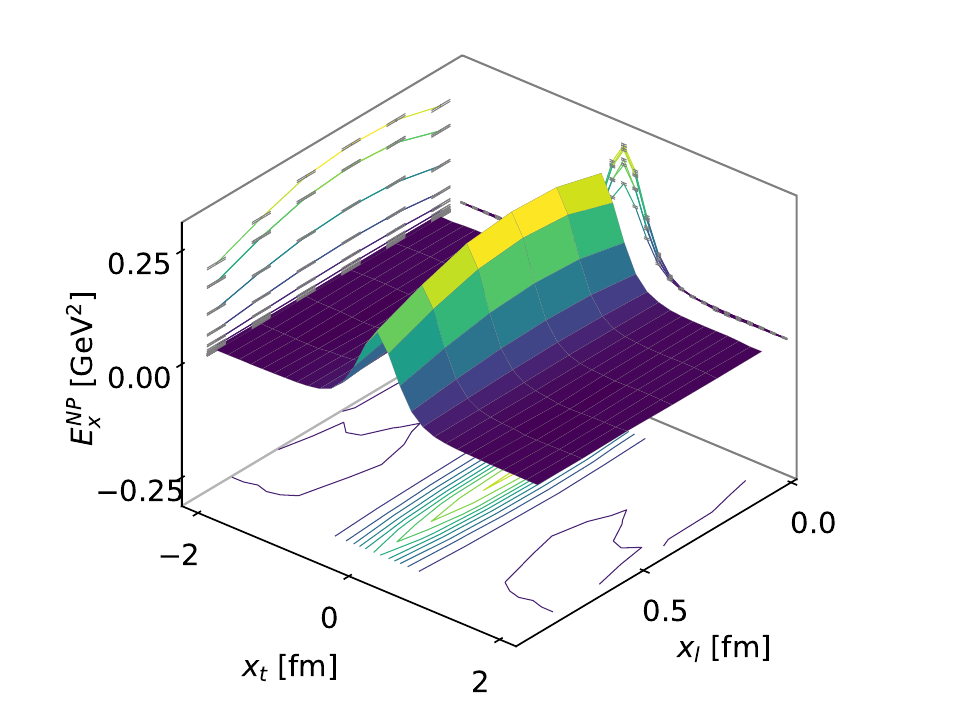}
\includegraphics[width=0.30\linewidth,clip]{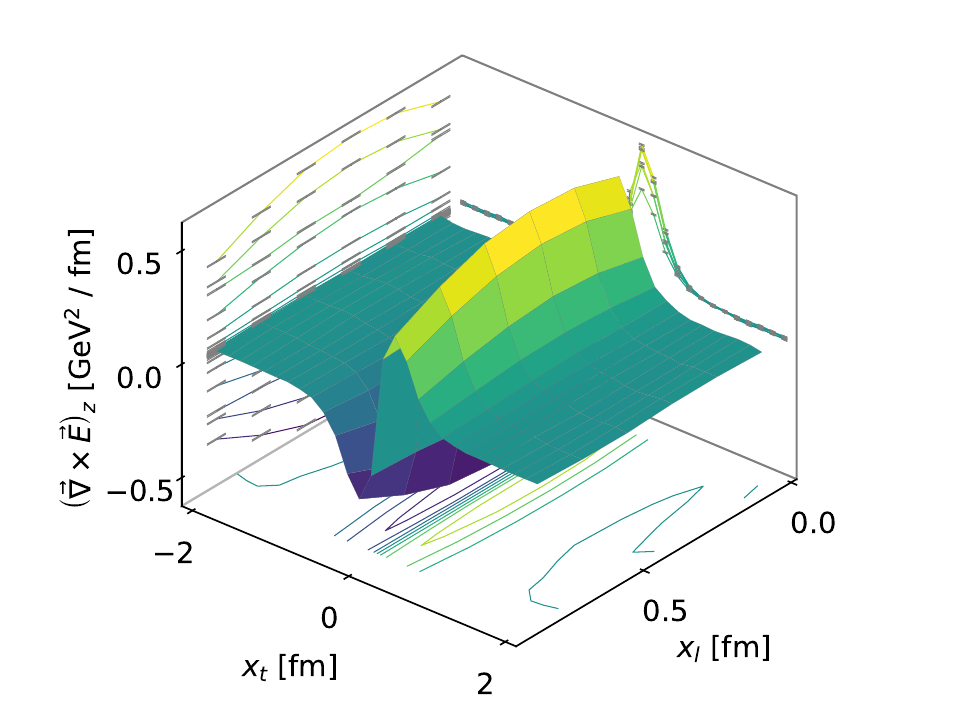}
\caption{Space distribution of the full electric field $E_x$ (left panel) and of its
nonperturbative part $E_x^{NP}$ (middle panel) for $\beta=6.3942$ and $d=6a=0.91$ fm.
The right panel displays the magnetic current at $\beta=6.3942$ and $d=0.91$ fm.}          
\label{fig:Ex_3d}
\end{figure*}
The connected correlator defined in \Cref{connected1} exhibits large
fluctuations at the scale of the lattice spacing, which are responsible
for a small signal-to-noise ratio. To extract the physical information carried
by fluctuations at the physical scale (and, therefore, at large distances
in lattice units), we smoothed out configurations by a {\em smearing}
procedure.

Our setup consisted of one step of 4-dimensional hypercubic smearing~\cite{Hasenfratz:2001hp} on the temporal links (HYPt), with smearing parameters $(\alpha_1,\alpha_2,\alpha_3) = (1.0, 1.0, 0.5)$, and $N_{\rm HYP3d}$ steps of hypercubic smearing  restricted to the three spatial directions (HYP3d) with $(\alpha_1^{\text{HYP3d}},\alpha_3^{\text{HYP3d}}) = (0.75, 0.3)$.

The operator in \Cref{connected1}, which defines the color field strength tensor, undergoes a  nontrivial renormalization~\cite{Battelli:2019lkz}, which depends on $x_t$. As discussed in \Onlinecite{Baker:2018mhw,Baker:2019gsi}, comparing our results with those in \Onlinecite{Battelli:2019lkz}, we argued that smearing behaves as an effective renormalization. In Fig.~\ref{fig:example_of_smearing} we show an example
of the behavior under smearing of the longitudinal component of the electric
field at the midplane between the sources and for different values of the transverse
position $x_t$: initially, smearing erases fluctuations at the level of the lattice spacing and leads to a rapid growth of the signal up to a maximum; thereafter, further smearing results in extended plateaus followed by a slow decrease, due to the fact that finally also the long-distance structure of the signal gets affected and degrades. We choose therefore the {\em optimal} number of smearing steps as the one for which the field takes its maximum value. It typically corresponds to a few units at small values of $x_t$ and to a few tens at larger values of $x_t$.
\begin {figure}[hbt]
\vspace{-0.01cm}
\centering
\includegraphics[width=\linewidth,clip]{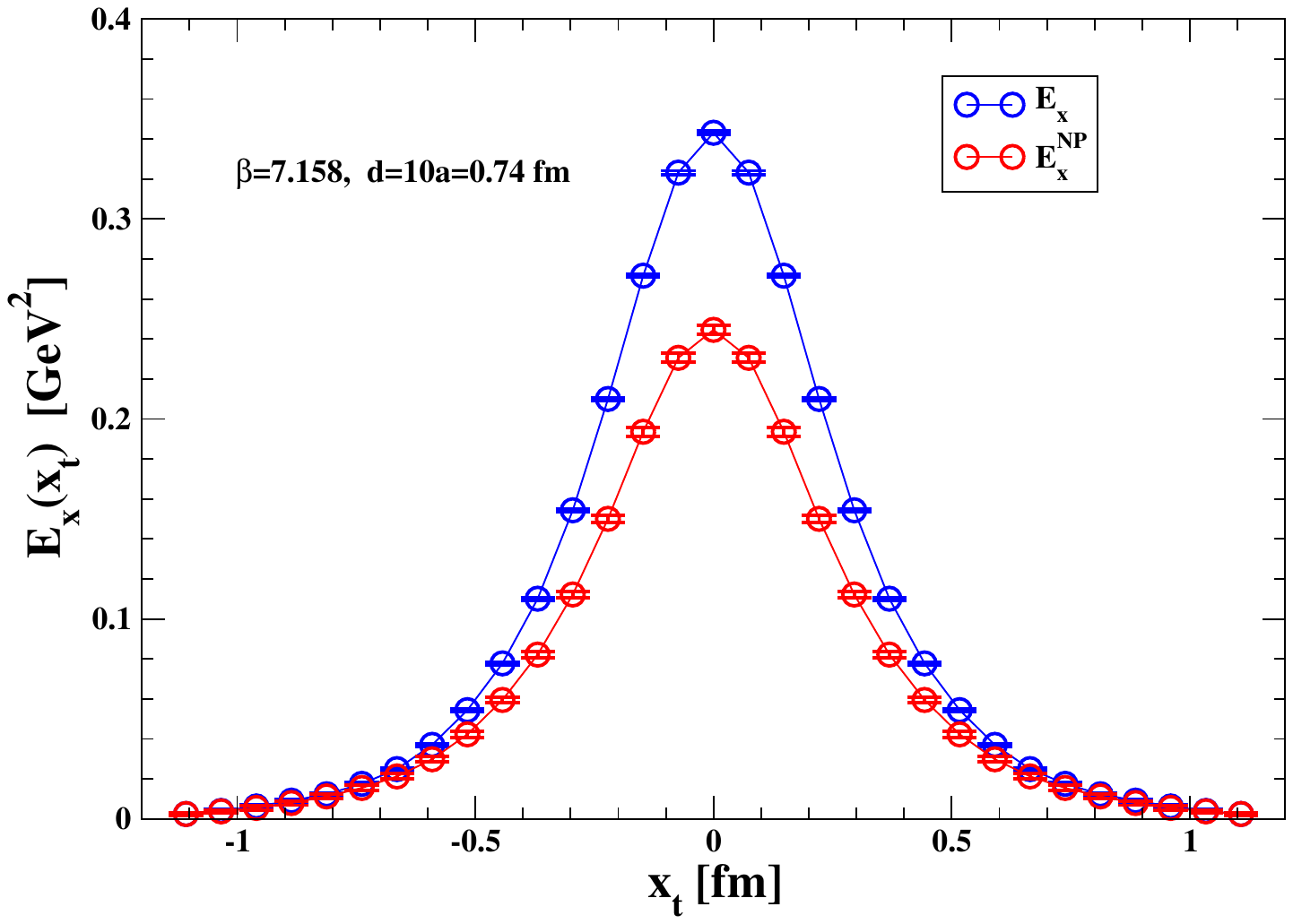}
\caption{Space distribution of the full electric field $E_x$ (left) and of its
nonperturbative part $E_x^{NP}$ (right) for $\beta=7.158$ and $d=10a=0.74\mathrm{fm}$ in correspondence of the midplane.}               
\label{fig:Ex_midplane}
\end{figure}

The smearing procedure can also be validated {\em a posteriori} by the observation of {\em continuum scaling}, {\it i.e.}, by checking that fields obtained in the same {\em physical} setup, but at different values of the coupling and of the optimal number of smearing steps, are in good agreement in the range of gauge couplings used (see subsection~\ref{scaling}).

In Table~\ref{numsimul} we display the list of measurements we have done. We considered distances between the sources in a wide interval, ranging from about 0.57~fm to about 1.37~fm, corresponding to $\beta$-values in the range 6.25765 to 7.158. 

\section{Numerical results}

\subsection{Scaling check}
\label{scaling}
We verified that our lattice setup is close enough to the 
continuum limit by checking that different choices of the lattice parameters, corresponding to the same physical distance $d$ between the sources, lead to the same values of the relevant observables when measured in physical units. To this purpose,
we considered the full electric field and its nonperturbative part at the midplane between the sources when the distance between the sources is in the range $d=0.723-0.760$ fm for six different lattice setups, in the range $d=0.855-0.959$ fm for six different lattice setups and in the range $d=1.013-1.060$ fm for four different lattice setups, see Fig.~\ref{fig:scaling}.
In all cases data points obtained on different lattices nicely overlap, taking into account the uncertainties and the spread of the $d$ values inside each 
of the three considered ranges. We notice from Fig.~\ref{fig:scaling}, middle and lower panels,   that the uncertainties become quite sizeable for the largest
value of $d$, especially in the nonperturbative part. This is a consequence of the degradation of the signal-to-noise ratio at larger distances, which, in the case of the nonperturbative field, is amplified by the curl subtraction.

\subsection{3d plots and asymmetry}
 We extract the nonperturbative part of the longitudinal electric field, $E_x^{NP}$, by systematic application of the {\em curl method}, first introduced in Ref.~\cite{Baker:2019gsi}, to which refer for the details of the procedure. In Fig.~\ref{fig:Ex_3d} we present two 3D plots giving the space distribution of the full longitudinal electric field $E_x$ (left panel) and of its nonperturbative part (middle panel): in the latter, a tube-like shape clearly emerges, fairly uniform in the longitudinal direction, up to a small asymmetry (the field at $x_l=0$ is a bit higher than the field at $x_l=d$) probably due to the asymmetric structure of the adopted lattice operator (see Fig.~\ref{fig:op_W}), which the smearing procedure is unable to balance. 
From data for the electric field, the magnetic current, given by the curl of the electric field, can also be extracted. For distances $d\lesssim 1$ it is a smooth function, see Fig.~\ref{fig:Ex_3d}, right panel, for an example.

In Fig.~\ref{fig:Ex_midplane} we show a transverse section of the electric field profile, comparing its full value and the nonperturbative part at the midplane between the sources: the latter is sensibly
smaller at $x_t=0$, {\it i.e.}, on the longitudinal axis, but is roughly as wide as the full one.

\subsection{Field integrals: string tension and width of the flux tube}
\begin{table*}[htb]
\begin{center} 
  \caption{Numerical results for $\sqrt{\sigma_\mathrm{eff}}$ and  $w$, as defined in Eqs.~(\ref{sigma_eff}) and~(\ref{width}).}
  \label{stringandwidth_NP}
\setlength{\tabcolsep}{12pt}
\begin{tabular}{cllclS[table-format=1.9]S[table-format=1.7]}
\toprule
lattice & $\beta=10/g^2$ & $a(\beta)$ [fm]  & $d$ [lattice units]   & $d$ [fm] & $\sqrt{\sigma_\mathrm{eff}}$ & $w$ \\ \midrule
$48^4$  &  6.885    &  0.0949777  &  6   &  0.569866  &  0.33245(21)  &   0.65(8)   \\
$32^4$  &  7.158    &  0.0738309  &  8   &  0.590647  &  0.40933(6)   &   0.474(26) \\
$24^4$  &  6.445    &  0.144692   &  5   &  0.723462  &  0.39977(8)   &   0.46(5)   \\
$32^4$  &  7.158    &  0.0738309  &  10  &  0.738309  &  0.38097(26)  &   0.50(6)   \\
$48^4$  &  6.885    &  0.0949777  &  8   &  0.75982   &  0.38470(19)  &   0.46(7)   \\
$32^4$  &  6.885    &  0.0949777  &  8   &  0.759821  &  0.38018(21)  &   0.46(6)   \\
3$2^4$  &  6.5824   &  0.126658   &  6   &  0.759947  &  0.38227(12)  &   0.30(12)  \\
$32^4$  &  6.3942   &  0.15203    &  5   &  0.760151  &  0.37420(7)   &   0.51(5)   \\
$32^4$  &  6.885    &  0.0949777  &  9   &  0.854799  &  0.3526(6)    &   0.71(19)  \\
$32^4$  &  6.25765  &  0.173715   &  5   &  0.868575  &  0.35853(12)  &   0.37(7)   \\
$32^4$  &  6.5824   &  0.126658   &  7   &  0.886605  &  0.3715(4)    &   0.74(31)  \\
$32^4$  &  6.3942   &  0.15203    &  6   &  0.912182  &  0.35455(32)  &   0.56(14)  \\
$48^4$  &  6.885    &  0.0949777  &  10  &  0.949777  &  0.3707(18)   &   0.53(24)  \\
$32^4$  &  7.158    &  0.0738309  &  13  &  0.959801  &  0.368(5)     &   0.61(27)  \\
$32^4$  &  6.25765  &  0.173715   &  6   &  1.04229   &  0.3369(8)    &   0.44(10)  \\
$32^4$  &  6.885    &  0.0949777  &  11  &  1.04475   &  0.293(6)     &   0.62(34)  \\
$32^4$  &  6.3942   &  0.15203    &  7   &  1.06421   &  0.3071(18)   &   0.50(18)  \\
\bottomrule 
\end{tabular}
\end{center}
\end{table*}
\begin{table*}[htb]
\begin{center} 
  \caption{Numerical results for $\sqrt{\sigma_\mathrm{eff}}$ and  $w$, as defined in Eqs.~(\ref{sigma_eff}) and~(\ref{width}) with the replacement of the nonperturbative field with the full one.}
  \label{stringandwidth}
\setlength{\tabcolsep}{12pt}
\begin{tabular}{cllclS[table-format=1.10]S[table-format=1.7]}
\toprule
lattice & $\beta=10/g^2$ & $a(\beta)$ [fm] & $d$ [lattice units]   & $d$ [fm] & $\sqrt{\sigma_\mathrm{eff}}$ & $w$ \\ \midrule
$48^4$  &  6.885        &  0.0949777    &  6    &  0.569866   &   0.48918(18)  &  0.62(7)    \\
$32^4$  &  7.158        &  0.0738309    &  8    &  0.590647   &   0.605219(33) &  0.447(18)  \\
$24^4$  &  6.445        &  0.144692     &  5    &  0.723462   &   0.53767(5)   &  0.436(35)  \\
$32^4$  &  7.158        &  0.0738309    &  10   &  0.738309   &   0.52234(13)  &  0.47(4)    \\
$48^4$  &  6.885        &  0.0949777    &  8    &  0.75982    &   0.51466(11)  &  0.45(5)    \\
$32^4$  &  6.885        &  0.0949777    &  8    &  0.759821   &   0.51557(10)  &  0.44(4)    \\
$32^4$  &  6.5824       &  0.126658     &  6    &  0.759947   &   0.52667(5)   &  0.36(8)    \\
$32^4$  &  6.3942       &  0.15203      &  5    &  0.760151   &   0.516572(32) &  0.459(26)  \\
$32^4$  &  6.885        &  0.0949777    &  9    &  0.854799   &   0.45568(28)  &  0.55(11)   \\
$32^4$  &  6.25765      &  0.173715     &  5    &  0.868575   &   0.47216(5)   &  0.34(4)    \\
$32^4$  &  6.5824       &  0.126658     &  7    &  0.886605   &   0.45257(17)  &  0.32(30)   \\
$32^4$  &  6.3942       &  0.15203      &  6    &  0.912182   &   0.46500(12)  &  0.16(5)    \\
$48^4$  &  6.885        &  0.0949777    &  10   &  0.949777   &   0.4273(10)   &  0.49(20)   \\
$32^4$  &  7.158        &  0.0738309    &  13   &  0.959801   &   0.4008(18)   &  0.50(15)   \\
$32^4$  &  6.25765      &  0.173715     &  6    &  1.04229    &   0.37883(27)  &  0.42(5)    \\
$32^4$  &  6.885        &  0.0949777    &  11   &  1.04475    &   0.3550(26)   &  0.58(20)   \\
$32^4$  &  6.3942       &  0.15203      &  7    &  1.06421    &   0.3759(6)    &  0.47(8)    \\
$32^4$  &  6.33727      &  0.160714     &  7    &  1.125      &   0.3151(27)   &  0.57(33)   \\
$32^4$  &  6.314762     &  0.164286     &  7    &  1.15       &   0.4258(35)   &  0.44(17)   \\
$24^4$  &  6.445        &  0.144692     &  8    &  1.157536   &   0.426(5)     &  0.60(27)   \\
\bottomrule 
\end{tabular}
\end{center}
\end{table*}
\begin {figure*}[htb]
\centering
\includegraphics[width=0.48\linewidth,clip]{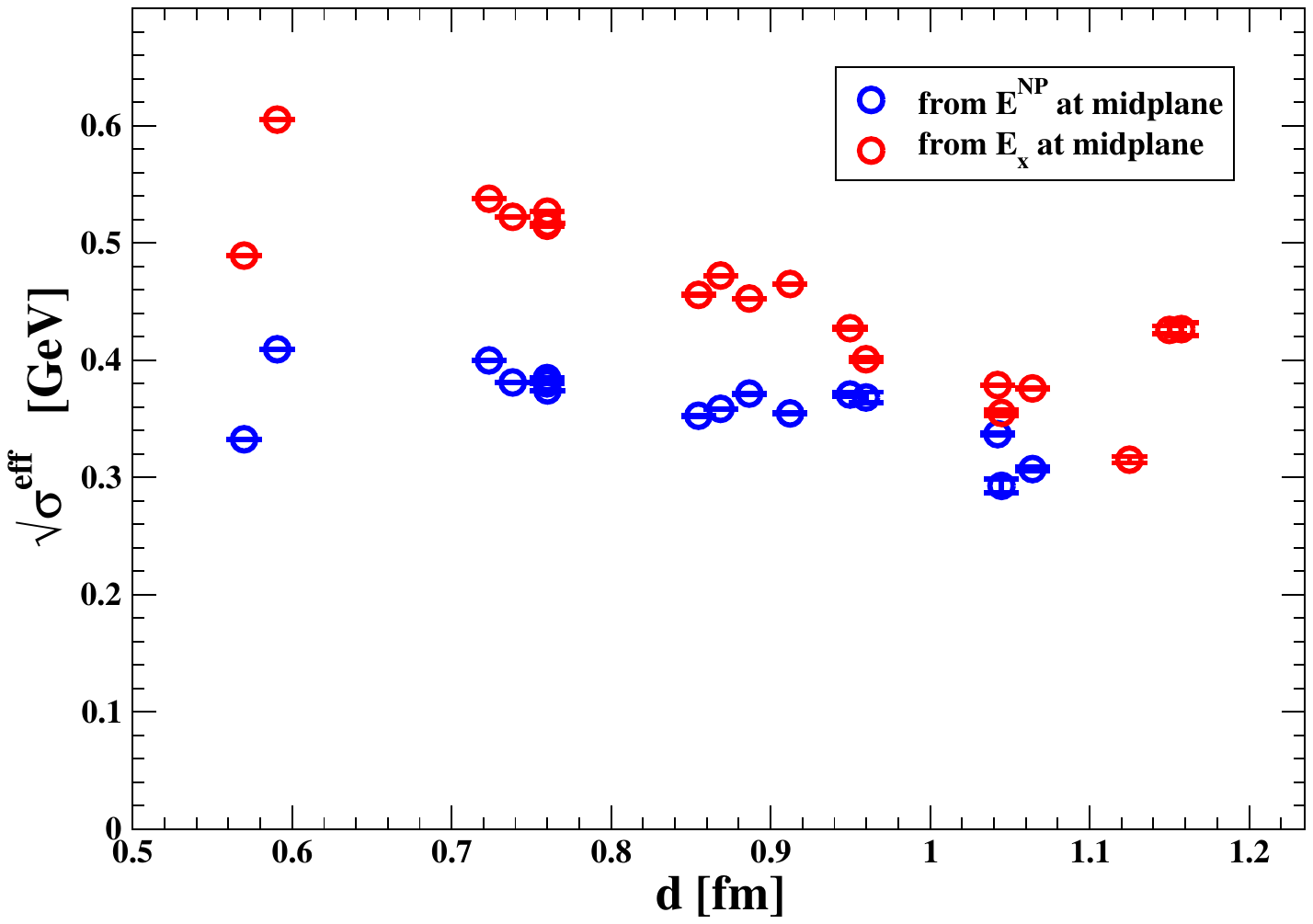}
\hspace{0.2cm}
\includegraphics[width=0.48\linewidth,clip]{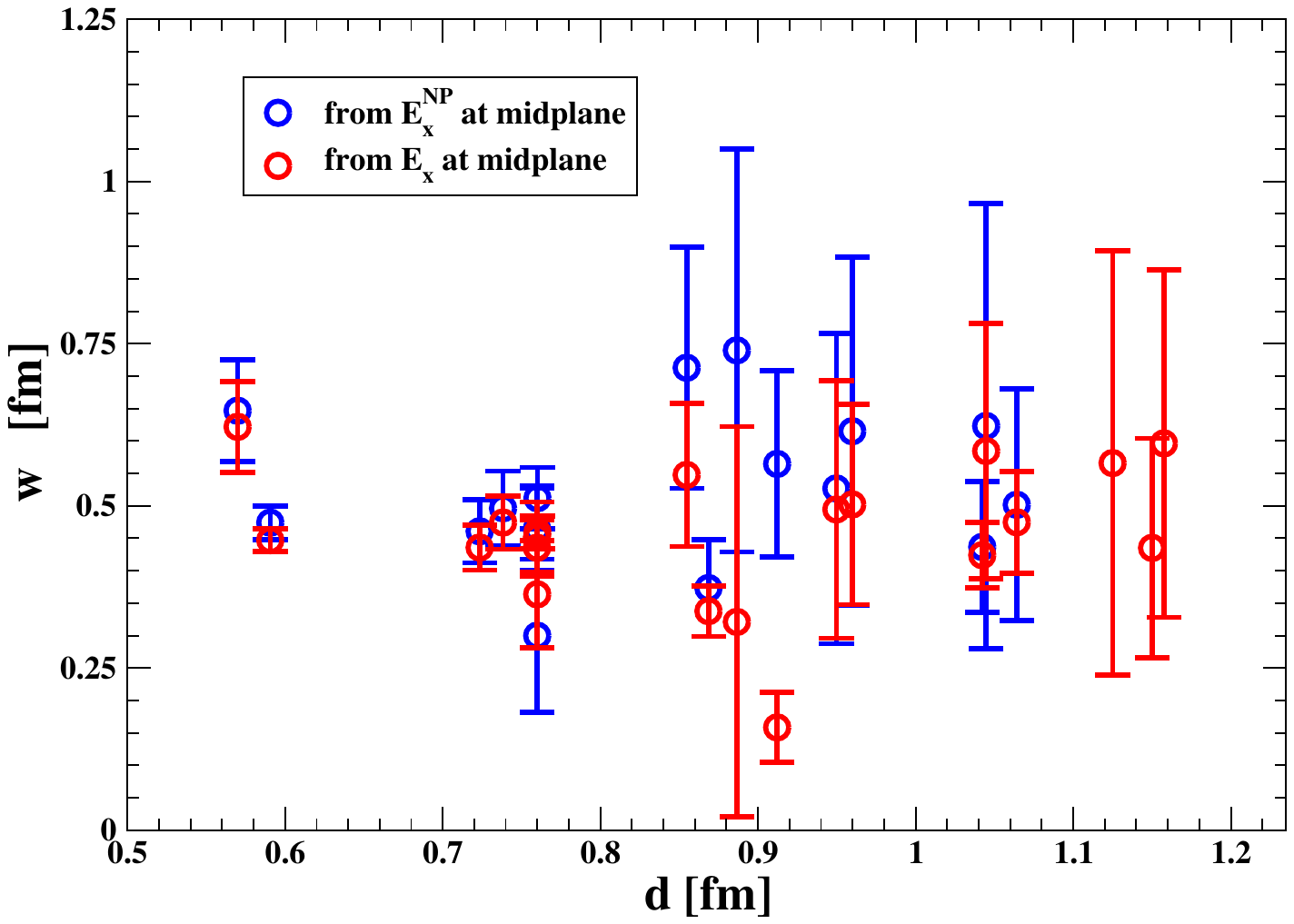}
\caption{ Behaviour of $\sigma_{\rm eff}$ (left panel)  and $w$ (right panel) 
with the distance $d$ between the sources, for the full longitudinal electric field (red circles) and its nonperturbative part (blue circles).}               
\label{fig:string}
\end{figure*}

To characterize quantitatively the shape and some properties
of the flux tube formed by the longitudinal electric field, we calculated
numerically the following two expressions, involving three different integrals,
\begin{equation}
\sigma_\mathrm{eff} = \int d^2 x_t \ \frac{(E_x^\mathrm{NP}(x_t))^2}{2} \ , 
\label{sigma_eff}
\end{equation}
\begin{equation} 
w = \sqrt{\frac{\int d^2x_t \, x_t^2 E_x^\mathrm{NP}(x_t)}{\int d^2x_t \, E_x^\mathrm{NP}(x_t)}} \ ,
\label{width}
\end{equation}
with the longitudinal electric field taken at the midplane between the sources.
The first of them represents a quantity which has the dimension of an energy per unit length, similarly to the string tension. However, this quantity, denoted by $\sigma_{\rm eff}$, does not coincide with the true string tension, for which the integrand would contain $\sum_{a=1}^8 (E_x^a)^2$. In (\ref{sigma_eff}) we have instead the squared Maxwell-like field, which is arguably a linear combination of the Abelian color components 3 and 8 of the electric field (see next Section). The expression in Eq.~(\ref{width}) gives an estimate of the width of the flux tube.

The integrals in Eqs.~(\ref{sigma_eff}), (\ref{width}) are computed numerically by means of the trapezoidal rule. 
They were considered both for the nonperturbative part of the longitudinal electric field and for the full field. The numerical results are
displayed, respectively, in Tables~\ref{stringandwidth_NP} and~\ref{stringandwidth}.

In Fig.~\ref{fig:string}, left panel,  we compare the behavior of $\sigma_{\rm eff}$ with the 
distance $d$ between the sources for the full longitudinal electric field
and its nonperturbative part: while for the full field $\sigma_{\rm eff}$ tends to decrease, for the nonperturbative part it is fairly stable. This different behavior is not surprising: the full field 
on the midplane contains also the perturbative contribution, which becomes less and less relevant when the distance between the sources increases.
While not visible on the figure, the uncertainties in the estimation of the string tension increase up to two orders, when the distance $d$ between sources increases, as can be seen in Tables~\ref{stringandwidth_NP} and~\ref{stringandwidth}.

In Fig.~\ref{fig:string}, right panel,  we show a similar comparison for the width. Differently from the case of $\sigma_{\rm eff}$, within uncertainties which increase with $d$, the width of the flux tube remains stable on a wide range of distances and is generally
compatible for the full and the nonperturbative field. The uncertainties in the estimation of the width are much larger than the ones for the estimation of the string tension, due to the fact that the width is defined as a ratio of two numerical integrals. 

From the constant fit of the data for string tension and string width we get $\sqrt{\sigma_{\rm eff}} \approx 0.4\ \mathrm{GeV}$, and 
$w \approx 0.5\ \mathrm{fm}$.

\section{Magnetic currents and electric fields}
\label{CM}
We have seen how the Maxwell picture of the flux tube leads to a well-defined mechanism for the squeezing of the electric fields into
a narrow flux tube. More precisely, the presence of a magnetic current density that circulates around the axis of the flux tube
ensures the squeezing of the electric flux tube in the transverse direction according to the Maxwell equation, Eq.~(\ref{rotel}),
that for later convenience we rewrite here as:
\begin{equation}
\label{CM1}
   \vec{\nabla} \; \times \vec{E}(\vec{x}) \; = \; \vec{J}_{\rm mag}(\vec{x})     \;  \; .
\end{equation}
Indeed, if in cylindrical coordinates:
\begin{equation}
\label{CM2}
\vec{J}_{\rm mag}(\vec{x})     \; \simeq  \; J_{\rm mag}(x_t) \; {\hat \phi }
\end{equation}
and the function  $J_{\rm mag}(x_t)$ decreases sufficiently fast toward zero when $x_t \, \rightarrow \, \infty$, then
\cref{CM1} assures the presence of an almost uniform and narrow longitudinal electric field along the flux tube.
Actually, the general solution of  \cref{CM1} is given by:
\begin{equation}
\label{CM3}
 \vec{E}(\vec{x}) \; = \; \vec{E}^{\rm NP}(\vec{x})     \;  + \;  \vec{E}_C(\vec{x}) \; \; ,
\end{equation}
where $\vec{E}_C(\vec{x})$ is the solution of the homogeneous equation   $\vec{\nabla} \; \times \vec{E}_C(\vec{x}) \, = \, 0$ and
 $\vec{E}^{\rm NP}(\vec{x})$ satisfies the  nonhomogeneous equation:
\begin{equation}
\label{CM4}
\vec{\nabla} \; \times \vec{E}^{\rm NP}(\vec{x}) \; = \; \vec{J}_{\rm mag}(\vec{x})     \;  \; .
\end{equation}
\begin{table*}[thb]
\begin{center}
\setlength{\tabcolsep}{20pt}
\begin{tabular}{cS[table-format=1.7]S[table-format=1.7]S[table-format=1.6]c}
\hline
 $x_{\ell}$ [fm] &   \multicolumn{1}{c}{$v_{\phi}$}  &  \multicolumn{1}{c}{$\sqrt{gH_0}$ [GeV]} &  \multicolumn{1}{c}{$\alpha$}  & $\chi^2_r$    \\
\hline
$0.076$  & 0.079(24) & 1.04(13)  & 0.82(17) &  1.98 \\
$0.228$  & 0.120(10) & 0.908(28) & 1.0      &  1.53 \\ 
$0.380$  & 0.22(11)  & 0.73(12)  & 1.4(6)   &  1.03 \\
$0.532$  & 0.20(6)   & 0.72(8)   & 1.4(4)   &  1.03 \\
$0.684$  & 0.13(5)   & 0.83(10)  & 1.02(28) &  0.19 \\
$0.836$  & 0.081(21) & 0.97(9)   & 0.81(15) &  1.22 \\
$0.988$  & 0.068(9)  & 0.94(5)   & 1.0      &  1.42 \\
\hline
\end{tabular}
\end{center}
\caption{Values of the parameters for the fits of the lattice data to \cref{CM9} along the flux tube at  $\beta$ = 6.3942, 
$a(\beta) \simeq 0.15203$ fm, corresponding to static color source distance $d$  = 7 a $\simeq$ 1.064  fm.}
\label{TableCM1}
\end{table*}
\begin {figure*}[thb]
\centering
\includegraphics[width=0.47\linewidth,clip]{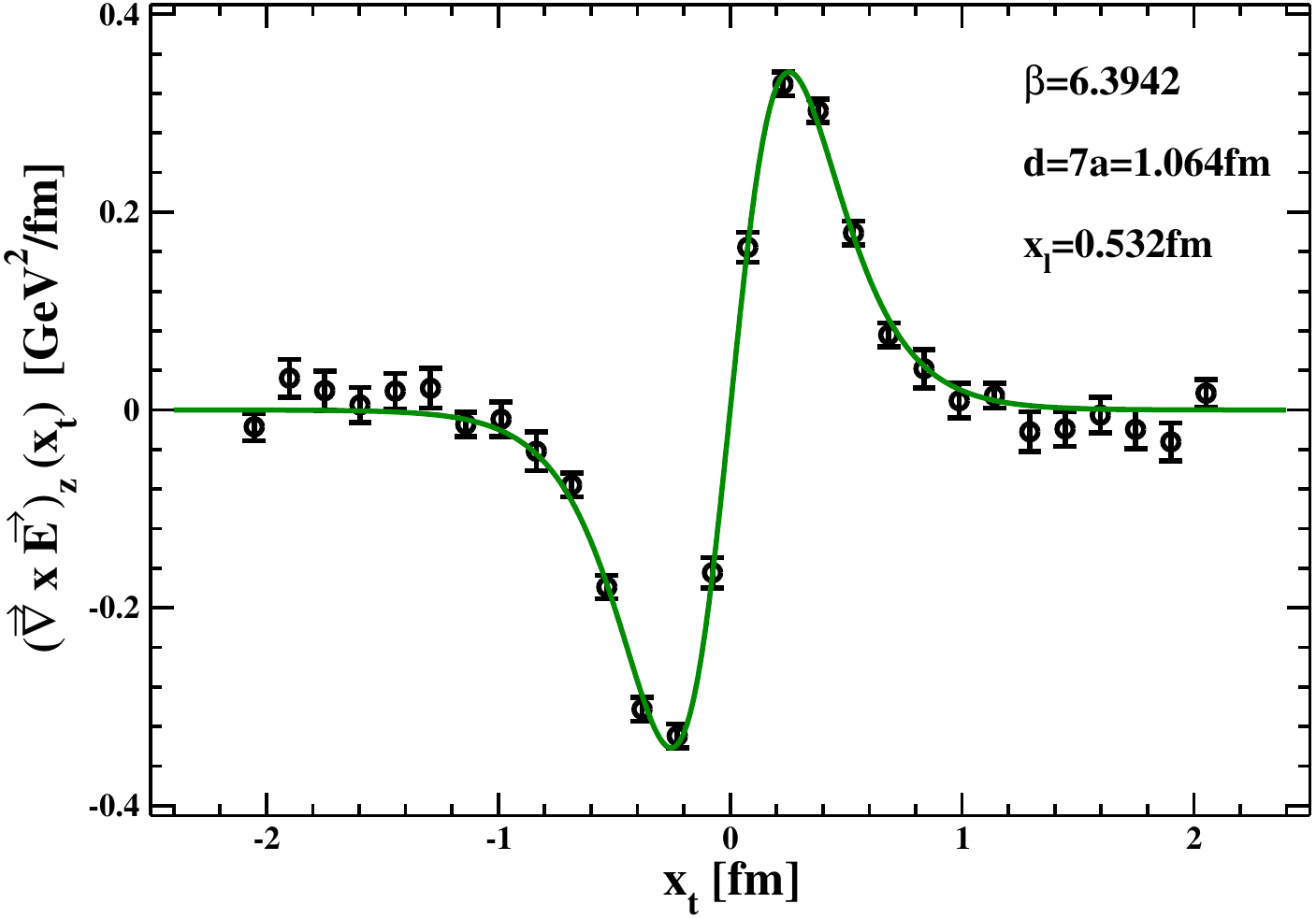} 
\hspace{0.3cm}
\includegraphics[width=0.47\linewidth,clip]{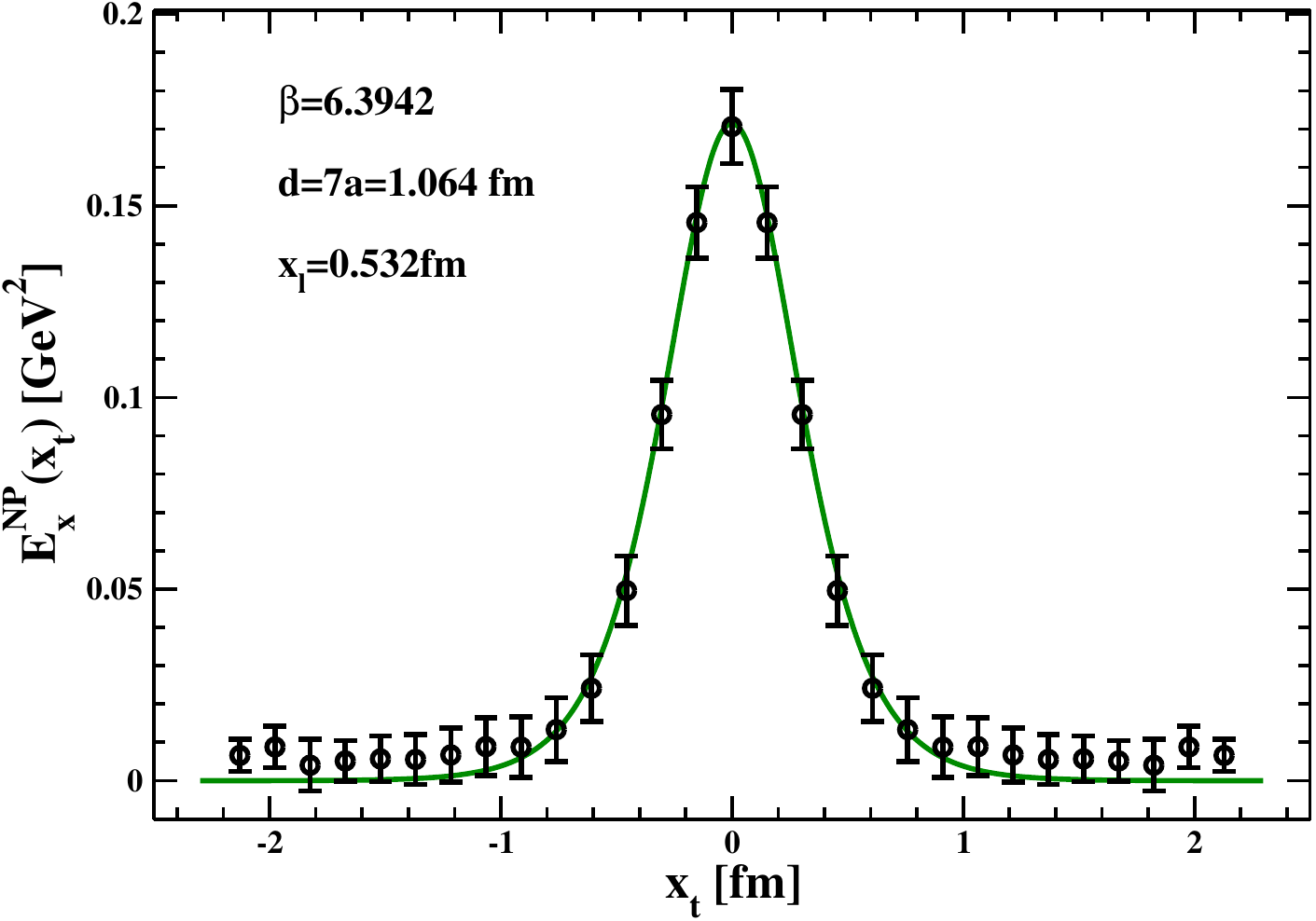}
\caption{
(Left panel) The lattice data for the curl of the electric field compared to the total magnetic current (full green line).
(Right panel) Comparison of the transverse distribution of the nonperturbative flux-tube electric field to $E_{th}(x_t)$, 
\cref{CM11} (full green line).}
\label{FigCM1}
\end{figure*}

This last equation shows that the nonperturbative electric fields  $\vec{E}^{NP}(\vec{x})$ turn out to be squeezed into a narrow
flux tube leading to a nonzero string tension. Our previous nonperturbative and model-independent analysis allowed to evaluate the
Coulomb field $\vec{E}_C(\vec{x})$ by solving numerically $\vec{\nabla} \; \times \vec{E}(\vec{x}) \, = \, 0$ and, after that, to determine
the nonperturbative electric field by means of \cref{CM3}. Moreover, by evaluating numerically the curl of the full electric
field, we were able to check the validity of \cref{CM2} and to track the transverse distribution of the magnetic current. 
To make further progress, we need some theoretical input on the structure of the magnetic currents. Recently, a first principle
attempt to characterize the structure of the QCD vacuum at large distances has been advanced in Ref.~\cite{Cea:2023}.
According to Ref.~\cite{Cea:2023}, the QCD vacuum resembles a disordered magnetic condensate with average strength
$\sqrt{gH_0} \, \sim 1.0$ GeV such that color confinement is assured by the presence of a mass gap together with the lack of
color long-range order. Interestingly enough, in Ref.~\cite{Cea:2023} there is a physical picture for the formation of the flux tube between
static  color sources. As a matter of fact, the presence of a static quark-antiquark pair leads to the polarization of the
magnetic domains characterizing the QCD vacuum, such that there are magnetic currents circulating around
the line joining the static color charges that, in turn, give rise to a Lorentz force that is  able to squeeze the electric fields
generated by the quark-antiquark pair. More importantly, it turned out that, within the approximations adopted in 
Ref.~\cite{Cea:2023} and sufficiently far from the color sources, the induced magnetic currents belong to the maximal
Abelian subgroup of the SU(3) gauge group. As a consequence, the equations relating the magnetic currents to the
flux-tube electric fields are the familiar Maxwell equations generalised to the case of  the presence of magnetic 
charges~\cite{jackson_classical_1999}:
\begin{equation}
\label{CM5}
 - \;  \vec{\nabla} \; \times \vec{E}^a(\vec{x}) \; = \; \vec{J}^{\,a}_M(\vec{x})     \;  \; .
\end{equation}
Far from the color sources one has:
\begin{equation}
\label{CM6}
  \vec{J}^{\,a}_{\rm M}(\vec{x})  \; \simeq \;  \delta^{a3} \,  \vec{J}^{\;3}_{\rm M}(\vec{x})  \; + \;    \delta^{a8} \, \vec{J}^{\;8}_{\rm M}(\vec{x}) \;,
\end{equation}
where $\vec{J}^{\;3,8}_{\rm M}(\vec{x})$ are azimuthal magnetic currents with strengths~\cite{Cea:2023}: 
\begin{equation}
\label{CM7}
  J^{3}_{\rm M}(\vec{x}) \;  \simeq \;  - \;  v_{\phi}  \, \sqrt{2} \, (gH_0)^{\frac{3}{2}} \;  \, \frac{ \tanh \left ( \sqrt{\frac{gH_0}{2}} \, x_t  \right ) }
 {\cosh^2  \left ( \sqrt{\frac{gH_0}{2}} \, x_t  \right )    }   \;  \; ,
\end{equation}
and
\begin{equation}
\label{CM8}
 J^{8}_{\rm M}(\vec{x}) \;  \simeq \;  - \;  v_{\phi}  \, \frac{\sqrt{3}}{2}  \, (gH_0)^{\frac{3}{2}} \;  \, 
 \frac{ \tanh \left ( \alpha \, \sqrt{\frac{gH_0}{4}} \, x_t  \right ) }
 {\cosh^2  \left ( \alpha \,\sqrt{\frac{gH_0}{4}} \, x_t  \right )    }    \;  \; .
\end{equation}
In \cref{CM7} and \cref{CM8}  $v_{\phi}$ is the azimuthal velocity of the polarised magnetic domains. \\
Note that the magnetic currents basically depend only on two parameters, {\it i.e.}, $v_{\phi}$ and $\sqrt{gH_0}$. However,
we have added one more parameter $\alpha$ that away from the static color sources should be  $\alpha \simeq 1$~\cite{Cea:2023}.
As we shall see later on, this new parameter will allow us to track the lattice data for the curl of the electric field even near the color sources.
According to  Ref.~\cite{Cea:2023} the total magnetic current is:
\begin{equation}
\label{CM9}
 J_{\rm mag}(x_t)  \; \simeq \;  - \,  J^3_{\rm M}(x_t)  \; - \;   J^8_{\rm M}(x_t)  \; \; .
\end{equation}
Having an explicit expression for the current allows us to find the confining electric field:
\begin{equation}
\label{CM10}
 E^{\rm NP}(x_t)  \;  \simeq \;  E_{\rm th}(x_t) \;,
\end{equation}
with
\begin{equation}
\label{CM11}
 E_{\rm th}(x_t)  \;  \simeq \;  \, E^3_{\rm th}(x_t) \; + \;   E^8_{\rm th}(x_t) \; \; ,
\end{equation}
where
\begin{equation}
\label{CM12}
  E^3_{\rm th}(x_t) \; \simeq \;   v_{\phi} \;   gH_0 \;  \, \frac{1}{\cosh^2  \left ( \sqrt{\frac{gH_0}{2}} \, x_t  \right )  }   
\end{equation}
and
\begin{equation}
\label{CM13}
  E^8_{\rm th}(x_t) \; \simeq \; \frac{\sqrt{3}}{2 \, \alpha} \,   v_{\phi} \;   gH_0 \;  \, \frac{1}{\cosh^2 
   \left ( \alpha \,\sqrt{\frac{gH_0}{4}} \, x_t  \right )  }    \;  \; .
\end{equation}
\begin {figure*}[thb]
\centering
\includegraphics[width=0.48\linewidth,clip]{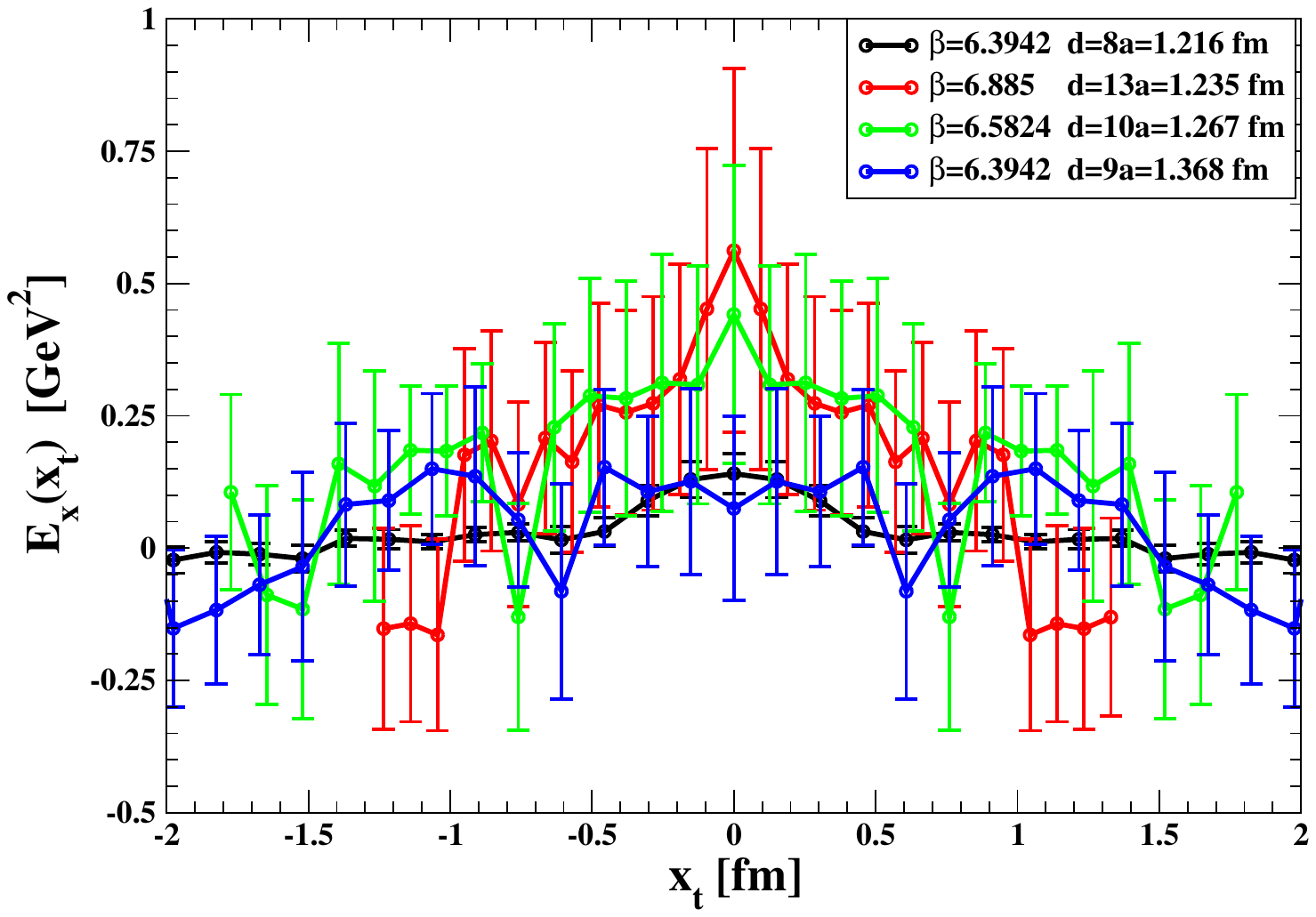}
\includegraphics[width=0.48\linewidth,clip]{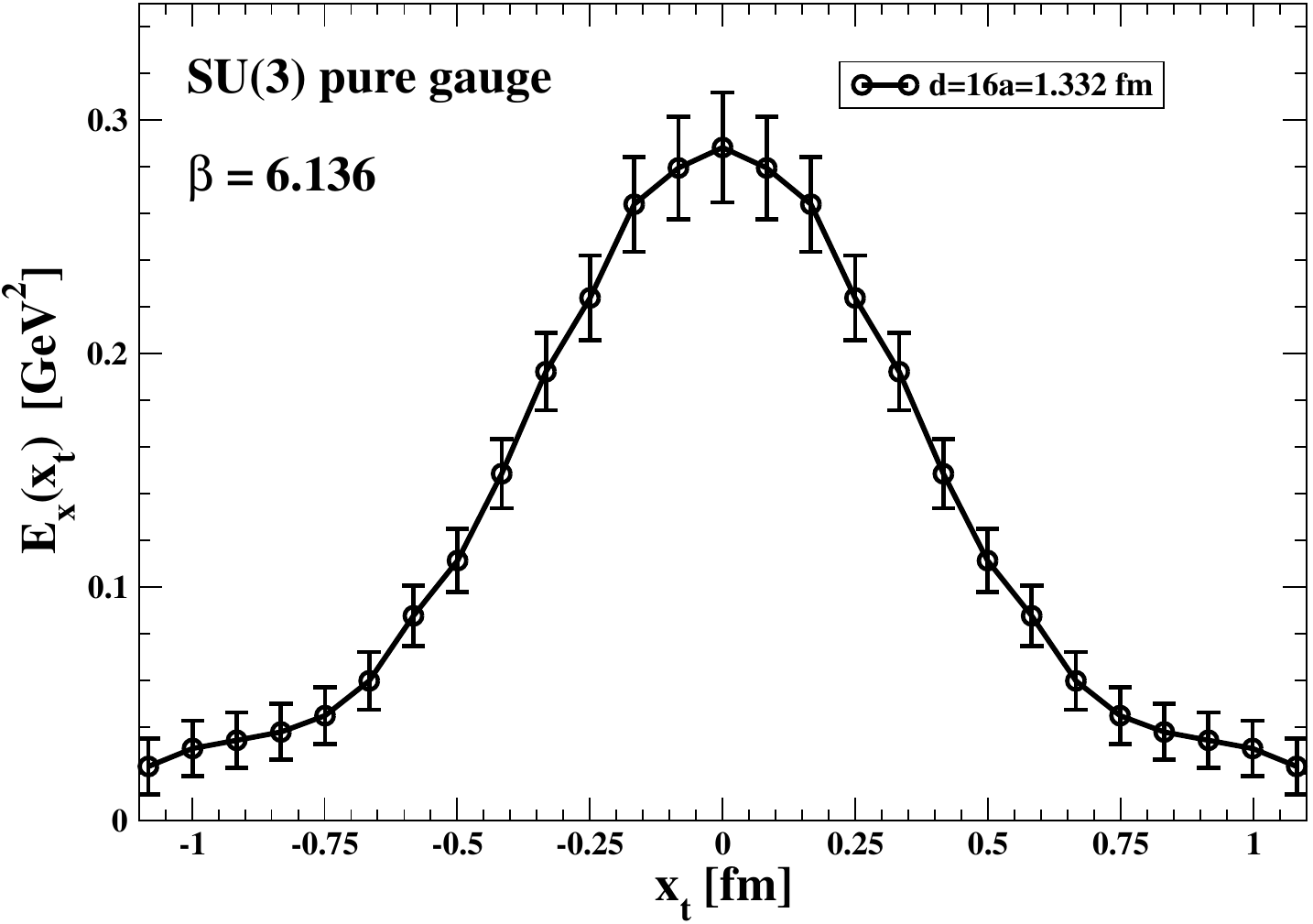}
\caption{Transverse profiles of the full longitudinal electric field on the midplane, for various distances between the sources.}       
\label{fig:stringbreaking}
\end{figure*}
A few comments are in order. Firstly, the Abelian nature of the confining electric fields supports our simplified Maxwell approach. The
total magnetic current, \cref{CM9}, depends on three parameters that can be fixed by performing a best fit to the lattice data. After that,
one can contrast \cref{CM11} to the lattice data. In this way we reach a nontrivial consistency check of our subtraction procedure to extract the 
nonperturbative electric field and, at the same time, of our Maxwell picture for confinement. Indeed, we have fitted our lattice data
for the curl of the electric field to \cref{CM9}. We found that, indeed, the total  magnetic current \cref{CM9} tracks quite well
the  lattice data for values of the distance between the quark-antiquark static pair up to about 1.1 fm (larger distances
will be discussed in the following Section). As an explanatory example, in  \cref{TableCM1} we report the best-fit parameters 
$v_{\phi}$,  $\sqrt{gH_0}$ and $\alpha$ along the flux tube corresponding to distance $ d \, \simeq \, 1.064$ fm. 
Also, for illustrative purpose, in \cref{FigCM1}, left panel,  we compare the best-fitted  magnetic current to the lattice data. \\
Once the parameters in the  magnetic current have been fixed by the best-fitting procedure,  we have a prediction for the  nonperturbative
confining  electric field, Eqs.~(\ref{CM11})-(\ref{CM13}). In fact, from the right panel in \cref{FigCM1} we see that 
 $E_{\rm th}(x_t)$ agrees almost perfectly with the data, confirming the consistency of our model-independent numerical procedure to extract
 the  nonperturbative field from the full flux-tube  electric field.
\section{Possible evidence for string breaking}
Although our numerical setup is not tailored for a clear-cut
detection of the expected {\em string breaking}, we tried to push our
numerical simulations to distances as large as $\sim 1.37$~fm, searching for hints of this phenomenon.
Indeed, in the presence of light quarks it is expected that the string between the static
quark-antiquark pair breaks at large distance  due to creation of a pair of light quarks
which recombine with the static quarks into two static-light mesons.
This expected zero-temperature phenomenon has proven elusive in simulations of lattice QCD and could not be clearly observed in early lattice simulations~\cite{Laermann:1998gm,Kratochvila:2003zj}.
Usually,   the string breaking distance $d^*$ is defined as the point where the
Wilson loop and the static-light meson operator have equal overlap onto the ground state,
{\it i.e.}, the mixing is most pronounced. Using this method, evidence for string breaking  was  first found in  Ref.~\cite{Bali:2005fu}
  for $N_f$ = 2  QCD on the lattice with pion mass about 640 MeV.
The authors of Ref.~\cite{Bali:2005fu} reported a string breaking distance
$d^*$  $\simeq$ 1.248(13)  fm. A more recent analysis employed 
$N_f$ = 2+1 flavors of  nonperturbatively improved dynamical Wilson fermions using an ensemble of gauge configurations generated
through the Coordinated Lattice Simulations (CLS) effort with  pion and kaon mass about   280 MeV and  460 MeV, 
respectively~\cite{Koch:2018puh}.
By employing  the Stochastic Laplacian Heaviside (LapH) method in order to efficiently calculate the correlation functions required for string breaking  and performing a variational analysis to extract the ground state as well as the first and second excited state of the system containing two static quarks, the authors of Ref.~\cite{Koch:2018puh} reported  a string breaking distance
$d^*$ $\approx$ 1.216 fm. Note that this last estimate is slightly smaller that the one in Ref.~\cite{Bali:2005fu}, as expected since 
the string breaking distance should decrease with the sea quark mass.\\
The main advantage of the present paper resides on the fact that we can look directly at the 
nonperturbative gauge-invariant longitudinal electric field, $\vec E^\text{NP}$, in the region between two static sources
that is responsible for the formation  of a well-defined flux tube, characterized by  nonzero  effective string tension $\sigma_{\rm eff}$ and  width $w$.
To this end, in Fig.~\ref{fig:stringbreaking}, left panel, we show the transverse profile of the full longitudinal  electric field on the midplane between two sources located at distances of 1.216~fm, 1.235~fm, 1.267~fm and 1.368~fm: at the latter distance the signal is almost completely lost. One might object that this is the effect of a degradation of the signal-to-noise ratio, due to the increasing distance {\em in lattice units}. This is however contradicted by the fact that at much larger values of the distance in lattice units, but a smaller physical distance, the signal is clearly present, see data for $d=1.235$~fm, corresponding to 13 lattice spacings, in Fig.~\ref{fig:stringbreaking}, left panel. \\
Even though for 1.216~fm $\lesssim  d < $1.368 fm we found evidence for the nonzero full longitudinal electric field on the midplane between two sources, we found no signs of a sizeable {\em nonperturbative}  longitudinal electric field.
This led us to extend the previous analysis to the smaller distances $d \simeq$ 1.140 fm, 1.157 fm,  1.183 fm (see Table~\ref{numsimul}).
Even in these cases, after subtracting the perturbative field from the full longitudinal electric field we found no clear evidence of a non-zero nonperturbative electric field.
In other words, for $d  \gtrsim 1.140$ fm we see no indication of the formation of the flux tube between the two color static sources. \\
Other facts that we observed are the following:
\begin{itemize} 
\item when the physical distance between quarks $d$ exceeds a certain value $d^*$, our method for the extraction of the longitudinal nonperturbative  electric field 
$E_x^{\rm NP}$ fails, due to the large uncertainties. Taking into account that we have a well-defined nonperturbative
flux tube up to $d \simeq 1.064$ fm (see the last entry in Table~\ref{stringandwidth_NP}), and above $d \simeq 1.140$ fm we cannot 
estimate the nonperturbative field, 
we can give a rough estimate of this distance as 1.064 fm $\lesssim d^* \lesssim$ 1.140 fm.
Note that our estimation seems to  be somewhat smaller with respect to Ref.~\cite{Koch:2018puh},
due to the fact that we are considering   QCD with (2+1) dynamical fermions at physical masses. 
Actually,  our determination of the string breaking distance is still slightly smaller with respect to  the recent estimate  extrapolated to the physical light-quark masses presented  in Ref.~\cite{Bulava:2024jpj}, but in good agreement with the estimate of the string breaking distance extrapolated to real QCD
of Ref.~~\cite{Bali:2005fu};
\item above $d^*$, no improvement can be observed if the distance in lattice units is reduced, keeping $d^*$ fixed; 
\item in SU(3) pure gauge theory by definition the string is unbroken. Here, at the physical distances $ d \sim d^*$ the signal is still there, see
 Fig.~\ref{fig:stringbreaking}, right panel.
\end{itemize}

In our approach, the string breaking distance $d^\ast$ should manifest in the drop of the (nonperturbative) electric field, and thus the decrease of the effective string tension $\sqrt{\sigma_\mathrm{eff}}$. Although our numerical data does not show a clear drop in $\sqrt{\sigma_\mathrm{eff}}$, the fact that above $d^\ast$ our method fails in the 2+1 flavour QCD, but not in the pure gauge theory, can itself be a sign of the string breaking.

From this analysis, we cannot make any firm claim about the onset of string breaking. We have just provided with some indirect hints of its appearance, which need to be corroborated by further tests.

\section{Conclusions}
We have investigated, by Monte Carlo numerical simulations of QCD with (2+1) dynamical staggered fermions at physical masses, the behavior of the nonperturbative gauge-invariant longitudinal electric field, $\vec E^\text{NP}$, in the region between two static sources, a quark and an antiquark.

We have performed our numerical simulations for a range of values of the coupling where continuum scaling is satisfied and have considered several values of the physical distance between the sources, ranging from about 0.5~fm up to
about 1.37~fm. Although our simulations allow us to depict the spatial distribution of all color fields in the region between the sources, for the 
sake of clarity and simplicity, most of the results presented in this paper
refer to the transverse plane midway between the sources.

Our findings can be summarized as follows. 

After subtraction of the perturbative component, the longitudinal  electric field takes the shape of a flux tube, whenever the distance between the sources
does not exceed a value $d^* \simeq 1.1$ fm. This flux tube can be characterized by two quantities, $\sigma_{\rm eff}$, which is related to the string tension, and the width $w$, which we have determined by numerical integration of our lattice data for the  electric field at the midplane between the sources.

Above the distance $d^*$, the longitudinal nonperturbative field is always compatible with zero, within large numerical uncertainties.
Our approach does not permit us to conclude if the disappearance of the nonperturbative signal at large distances is due to a signal-to-noise degradation or if it is instead a manifestation of the expected phenomenon of {\em string breaking}. We have provided some numerical arguments in favor of the latter hypothesis, but plan to corroborate them with future investigations.

We have compared our findings with a model of the QCD vacuum as disordered  magnetic condensate, finding a good compatibility between predictions and numerical results.

We plan to extend this analysis to the case of QCD with (2+1) flavors at nonzero temperature and density.

\section*{Acknowledgements}
This investigation was in part based on the MILC collaboration's public lattice gauge theory code (\url{https://github.com/milc-qcd/}). Numerical calculations have been made possible through a CINECA-INFN agreement, providing access to HPC resources at CINECA. PC, LC and AP acknowledge support from INFN/NPQCD project. VC acknowledges support by  the Deutsche Forschungsgemeinschaft \linebreak (DFG, German Research Foundation) through the CRC-TR 211  ``Strong-interaction matter under extreme conditions'' -- \linebreak project number 315477589 -- TRR 211. This work is (partially) supported by ICSC – Centro Nazionale di Ricerca in High Performance Computing, Big Data and Quantum Computing, funded by European Union – NextGenerationEU.

\bibliography{qcd}

\end{document}